\documentclass{aa}

\newcommand{\iso}       {\mbox{$ \rm sech^{2} $}}
\newcommand{\Rmax}      {\mbox{$  R_{\rm max} $}}
\newcommand{\Rmaxm}     {\mbox{$ \overline{R}_{\rm max} $}}
\newcommand{\zo}        {\mbox{$  z_{0}       $}}

\newcommand{\A}         {\mbox{$ ^{\rm a)}    $}}
\newcommand{\B}         {\mbox{$ ^{\rm b)}    $}}

\usepackage{graphicx}
\graphicspath{{/home/schwarz/bochum/papers/aa/99_2/figs/}{/ps/}}

\begin{document}

   \thesaurus{03         % A&A Section 11: Galaxies
              (11.05.2;  % Galaxies: evolution
               11.07.1;  % Galaxies: general,
               11.09.2;  % Galaxies: interactions,
               11.11.1;  % Galaxies: kinematics and dynamics,
               11.16.1;  % Galaxies: photometry,
               13.09.1)} % Infrared: galaxies.

   \title{The influence of interactions and minor mergers \\
          on the structure of galactic disks
          \thanks{Based on observations obtained at the European Southern Observatory
          (ESO, La Silla, Chile), Calar Alto Observatory operated by the MPIA (DSAZ, Spain),
          Lowell Observatory (Flagstaff/AZ, USA), and Hoher List Observatory (Germany).}
         }

   \subtitle{II. Results and interpretations}

   \authorrunning{U. Schwarzkopf \& R.-J. Dettmar}

   \titlerunning{The influence of interactions and minor mergers on disk structure. II}

   \author{U. Schwarzkopf \hspace{-0.6mm} \inst{1,}\inst{2,}\thanks{\emph{Present address:}
           Steward Observatory, University of Arizona,
           933 N. Cherry Ave., Tucson, Arizona 85721, U.S.A.}
          \and
           R.-J. Dettmar \inst{1}}

   \offprints{U. Schwarzkopf, \\ e-mail: schwarz@as.arizona.edu}

   \institute{Astronomisches Institut, Ruhr-Universit\"at Bochum,
              Universit\"atsstra{\ss}e 150, 44780 Bochum, Germany
             \and
              Steward Observatory, University of Arizona,
              933 N. Cherry Ave., Tucson, Arizona 85721, U.S.A.}

   \date{Received 9 September 1999 / Accepted 15 June 2000}

   \maketitle

   \begin{abstract}

We present the second part of a detailed statistical study focussed on the effects of tidal
interactions and minor mergers on the radial and vertical disk structure of spiral galaxies.
In the first part we reported on the sample selection, observations, and applied disk models.
In this paper the results are presented, based on disk parameters derived from a sample of
110 highly-inclined/edge-on galaxies. This sample consists of two subsamples of 49
interacting/merging and 61 non-interacting galaxies. Additionally, 41 of these galaxies
were observed in the NIR. We find significant changes of the disk structure in vertical
direction, resulting in $\approx 1.5$ times larger scale heights and thus vertical velocity
dispersions. The radial disk structure, characterized by the cut-off radius and the scale
length, shows no statistically significant changes. Thus, the ratio of radial to vertical
scale parameters, $h/\zo$, is $\approx 1.7$ times smaller for the sample of interacting/merging
galaxies. The total lack of typical flat disk ratios $h/\zo > 7$ in the latter sample implies
that vertical disk heating is most efficient for (extremely) thin disks. Statistically nearly
all galactic disks in the sample (93\%) possess non-isothermal vertical luminosity profiles like
the sech (60\%) and exp (33\%) distribution, independent of the sample and passband investigated.
This indicates that, in spite of tidal perturbations and disk thickening, the initial vertical
distribution of disk stars is not destroyed by interactions or minor mergers.

   \keywords{galaxies: evolution, general, interactions,
             kinematics and dynamics, photometry -- infrared: galaxies}

   \end{abstract}

%
%________________________________________________________________

\section{Introduction}

Although major galaxy mergers seem to be more spectacular and have therefore received
the most attention, there is evidence that tidal interactions or the accretion of small,
low-mass satellites (``minor mergers'') occur more frequently in the local universe
(e.g. Frenk et al. \cite{frenk1988}; Carlberg \& Couchman \cite{carlberg1989}).
%the more common process in the current epoch
%is the accretion of small, low-mass satellites -- hence ``minor mergers''
%(e.g. Frenk et al. \cite{frenk1988}; Carlberg \& Couchman \cite{carlberg1989}).
Considering the high density of galaxies in groups and clusters and the fact that a large
fraction of their members belongs to the dwarf galaxy population it is not unexpected
that we know a large number of galaxies influenced by environmental effects of this
order of magnitude. That includes both galaxies with clear signs of tidal interactions
that happend in the recent past and those currently involved merging processes
(Arp \cite{arp1966}; Arp \& Madore \cite{arp1987}; Fried \cite{fried1988}; Zaritsky et al.
\cite{zaritsky1993}, \cite{zaritsky1997}). A good example is our own Galaxy, that forms
a future minor merger together with the Large Magellanic Cloud
($M_{\rm LMC} \approx 2 \times 10^{10}~M_{\odot}$; $M_{\rm LMC}/M_{\rm MW} \approx 0.1$)
and with several other satellites (e.g. Sgr dwarf elliptical) having smaller mass ratios
(Irwin et al. \cite{irwin1985}; Schommer et al. \cite{schommer1992}; Toth \& Ostriker
\cite{toth1992}; Ibata \& Lewis \cite{ibata1998}). However, little is known about the
rate of minor merging events, their influence on the structure and kinematics of galactic
disks, and the efficiency of evoked perturbations.

In recent years this problem has been addressed by several numerical N-body simulations as
well as analytical estimations. It was found that minor mergers and accretion events in
the range $M_{\rm sat}/M_{\rm disk} \approx 0.05 \ldots 0.2$ must be a more common processes
in the local universe than previously argued, representing an important mechanism for driving
the evolution of galaxies. In particular, it was concluded that one of the most striking structural
changes produced by a single merger is a vertical disk thickening by a factor of 2--4, dependending
noticeably on the initial disk properties such as the ratio between scale length and -height $h/\zo$
(Ostriker \cite{ostriker1990}; Toth \& Ostriker \cite{toth1992}; Quinn et al. \cite{quinn1993};
Mihos et al. \cite{mihos1995}; Walker et al. \cite{walker1996}). However, the self-consistent
simulations recently carried out by Velazquez \& White (\cite{velazquez1999}) indicate that
these results tend to overestimate the vertical disk heating. They find that the heating
factor is closer to 1.5--2, depending mainly on the satellite orbit (i.e. prograde or retrograde)
and the mass of the bulge component. Furthermore, the obtained results might still be influenced
by counteracting processes such as dissipative gas cooling, subsequent star formation, the presence
of several stellar disk components, etc.
On the other hand, the observed thinness of typical late-type galaxy disks without indications
of tidal interaction/accretion constrains the value of vertical disk heating. It is thus unlikely
that such ``superthin'' galaxies have absorbed more than a few percent of their mass within their
lifetime (Toth \& Ostriker \cite{toth1992}).

At present there are only a few observational studies that aim at proving the effects predicted
by the simulations. Zaritsky (\cite{zaritsky1995}) analyzed observations of nearby spiral
galaxies based on magnitude residuals from the Tully-Fisher relationship, chemical abundance
gradients, and asymmetries in their stellar disks. He concluded that even relatively isolated
spiral galaxies have experienced accretion of companion galaxies over the last few Gigayears.
In their series of studies Reshetnikov et al. (\cite{reshetnikov1993}) and Reshetnikov \& Combes
(\cite{reshetnikov1996}, \cite{reshetnikov1997}) investigated the effects of tidally-triggered
disk thickening between galaxies of comparable mass. They used optical photometric data of a
sample of 24 interacting/merging and a control sample of 7 non-interacting disk galaxies.
As a main result they find that the ratio $h/\zo$ of the radial exponential disk scale length
$h$ to the constant scale height $\zo$ is about two times smaller for interacting galaxies.
However, the relatively small galaxy samples used in these studies make it difficult to derive
reliable estimates on the actual size of the structural changes. This also prevents a
consistent check with the results from simulations.

%Besides, the conclusions on their universality of the
%However, the main problem of most observational studies that have been undertaken so far was
%that they based on relatively small galaxy samples. Thus it can not reliably be concluded
%whether the observed changes of the disk structure due to interactions or minor mergers are
%consistent with simulation results.

Therefore we started a detailed statistical study in order to investigate systematically the
influence of interactions and minor mergers on the radial and vertical disk structure of spiral
galaxies in both optical and near infrared (NIR) passbands. Our study is based on a sample of 110
highly-inclined/edge-on disk galaxies, consisting of two subsamples of 61 non-interacting
galaxies and 49 interacting/minor merger candidates. Additionally, 41 of these galaxies were
observed in the NIR.

In Paper~I (Schwarzkopf \& Dettmar \cite{schwarzkopf2000a}) a detailed description of the
project structure and its main questions was given. We reported on the sample selection,
observations, and data reduction as well as on the disk modelling- and fitting procedure.

In Sect.~2 of this paper (Paper~II) the sample and applied corrections are briefly
summarized. In Sect.~3 we analyze the radial and vertical disk structure of both subsamples.
The global disk parameters, their ratios, and the vertical brightness distribution are
investigated. The derived colour gradients are also analyzed. We discuss the obtained
results in Sect.~4 and summarize and conclude the paper in Sect.~5.

%__________________________________________________________________

\section{The data}

\subsection{Sample and observations}

In Paper~I we found that it is crucial for this study to have two subsamples of
highly-inclined/edge-on galaxies ($i \ge 85 \degr$) that were selected carefully in
order to diminish overlapping effects, i.e. a contamination introduced by an uncertain
allocation of galaxies to the non-interacting or interacting/merging sample.
For the latter sample we therefore used a classification scheme that was introduced by
Arp \& Madore (\cite{arp1987}). Additionally, for most of the minor merger candidates
the mass ratio between the companion and the main body was checked. Finally, we have
shown that the distribution of the morphological types between both subsamples is
statistically indistinguishable over the whole range studied, i.e. between
$0 \le T \le 9$ (Paper~I). Although we cannot exclude overlapping effects completely,
the remaining uncertainties in the classification of the subsamples were thus reduced
to a minimum. A wrong allocation of objects would only lead to an underestimation of
the actual differences between both samples.

Since a large sample of galaxies was needed for this study the observations were obtained
with different telescopes and during several observing runs between February 1996 and
June 1998. Details of the observations and the data reduction can also be found in Paper~I.

%Many of the disk components investigated, preferably those of the intarcting/merging
%galaxy sample, show interesting and unusual disk,- bar- or bulge features, most probably
%evoked by the interacting process itself. Therefore, also some remarks on disk profiles
%of individual objects were given in Paper I.

%__________________________________________________________________

\subsection{Distances and corrections}

The distances to the observed galaxies with known redshifts were calculated using a
Hubble flow with a Hubble constant of $H_{0} = 75 \, \rm km \, \rm s^{-1} Mpc^{-1}$,
corrected for the ``Virgocentric Flow'' model predicted by Kraan-Korteweg
(\cite{kraan-korteweg1986}). The model chosen is characterized by a local infall
motion of the Local Group towards the Virgo cluster with an adopted velocity of
$v_{\rm vc} = 220 \, \rm km \, s^{-1}$.
It describes the motions of galaxies in the environment of the cluster by a
non-linear flow-model. Assuming this model and the adopted Hubble constant, the
distance of the Virgo cluster is $r_{\rm vir} = 15.8$~Mpc.

%%%%%%%%%%%%%%%%%%%%%%%%%%%%%%%%%%%%%%%%%%%%%%%%%%%%%%%%%%%%%%%%%%%

\begin{figure*}[t]

\vspace*{74mm}

\begin{minipage}[b]{8.8cm}
\begin{picture}(8.8,8.2)
{\includegraphics[angle=90,viewport=320 430 570 720,clip,width=86mm]{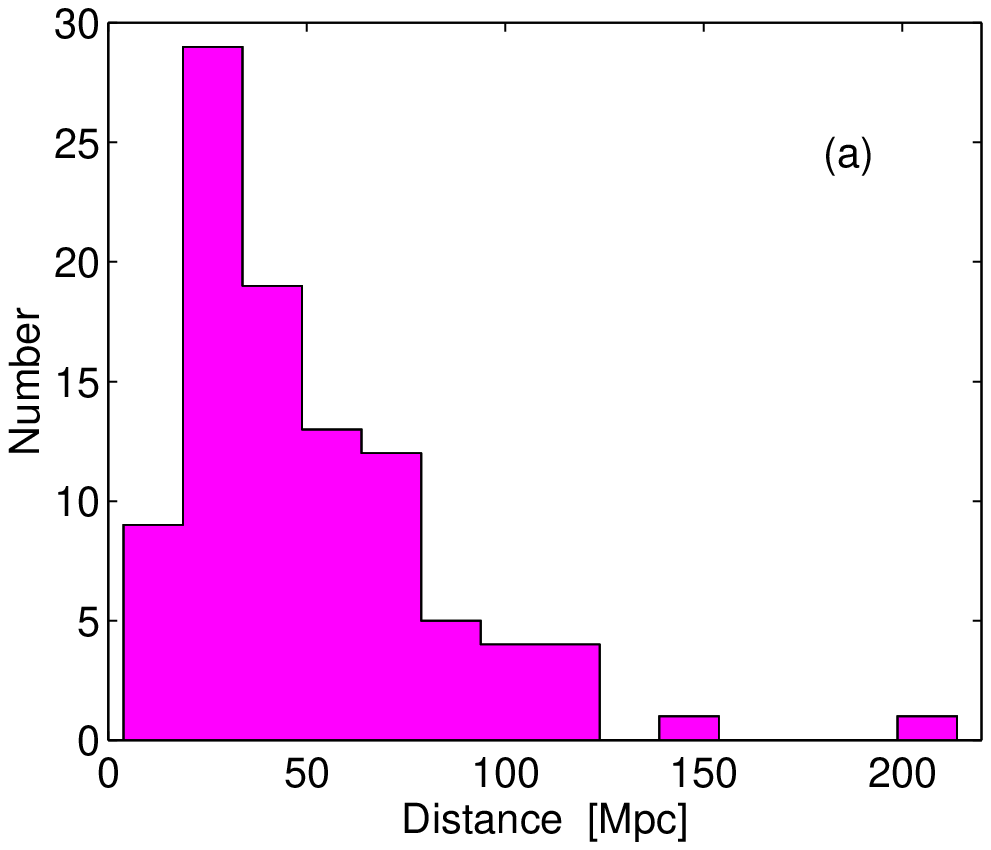}}
\end{picture} %landscape: angle =90; 320 430 570 720  portrait: angle=0; 40 210 270 400
\end{minipage}
\hfill
\begin{minipage}[b]{8.8cm}
\begin{picture}(8.8,7.7)
{\includegraphics[angle=90,viewport=320 430 570 720,clip,width=86mm]{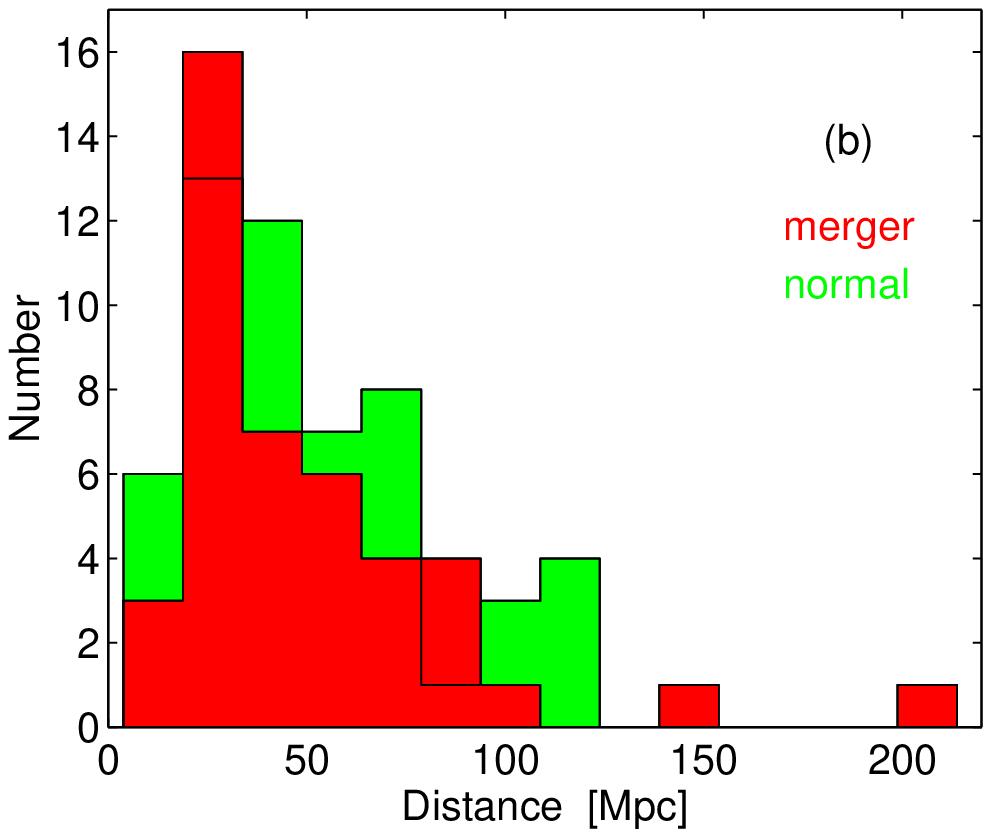}}
\end{picture}
\end{minipage}

\vspace*{0mm}

{\bf \noindent Fig.~1a and b.} The distribution of distances {\bf(a)} for the total
galaxy sample, and {\bf(b)} for both subsamples of non-interacting (normal) and
interacting/merging galaxies (all distances are corrected for Virgo infall).

\end{figure*}

%%%%%%%%%%%%%%%%%%%%%%%%%%%%%%%%%%%%%%%%%%%%%%%%%%%%%%%%%%%%%%%%%%%

The heliocentric velocities $v_{0}$ needed for this model are listed in
Table~\ref{parameters}, column~(5). They were calculated from optical/HI-velocities
taken from NED\footnote{The NASA/IPAC Extragalactic Database (NED) is operated by the
Jet Propulsion Laboratory, California Institute of Technology, under contract with the
National Aeronautics and Space Administration.}.
The heliocentric velocities $v_{0}$ were corrected for the system velocity
$v_{\rm LG}$ of the Local Group via formula (2) in Richter et al. (\cite{richter1987}).
The velocity corrections $\Delta v$ required for this model were derived from Fig.~3a
in Kraan-Korteweg (\cite{kraan-korteweg1986}).

The distance distribution of all sample galaxies with known redshifts ($n=97$) is
shown in Fig.~1a. The nearest galaxy is NGC 4244 at 3.8~Mpc, the most distant galaxy
is ESO 379-G20 at 193~Mpc (Table~\ref{parameters}).
Fig.~1b shows the distance distributions for both subsamples of 43 interacting/merging
and 54 non-interacting galaxies. According to the test of Kolmogorov \& Smirnov
(Darling \cite{darling1957}; Sachs \cite{sachs1992}) -- hereafter KS -- both
distributions are statistically indistinguishable:
the test result of 0.09 is significantly lower than the value necessary for the
20\%-limit (0.22), which is the strongest of the KS-criteria.
This is -- in addition to the indistinguishable distribution of morphological types
(Paper~I, Fig.~1) -- essential in order to avoid selection biases and thus to derive
reliable disk parameters.

Since seeing effects become significant for small disk structures such as the scale
height of flat disks (for seeing conditions around $2\arcsec$ and for features
$\le 4\arcsec$ the error amounts to 20\%) all disk parameters listed in
Table~\ref{parameters} were corrected in the following way:
using the values of predominant seeing conditions (derived from averaged FWHM values
of stars in the resulting frames, given in Paper~I, Table~2) the images were de-convoluted
with the Lucy-Richardson algorithm (standard MIDAS routine). In order to profit from the
full vertical resolution -- important to distinguish between different vertical disk models
used for fitting; Paper~I, Sect.~4) -- no vertical binning was applied.

%__________________________________________________________________

\subsection{Disk models and fitting procedure}

In order to analyze and to compare the structure of disk components of a large sample of
highly-inclined/edge-on spiral galaxies we developed an improved disk modelling- and
fitting procedure that is based on a 3-dimensional luminosity distribution proposed by
van der Kruit \& Searle (\cite{kruit1981a},\cite{kruit1981b}; \cite{kruit1982a}),
hereafter KS I--III. The results presented in the next Section were derived using
these disk models. An detailed description of their properties and the determination
of global disk parameters was given in Paper~I.

%__________________________________________________________________

%%%%%%%%%%%%%%%%%%%%%%%%%%%%%%%%%%%%%%%%%%%%%%%%%%%%%%%%%%%%%%%%%%%

\begin{figure*}[t]

\vspace*{74mm}

\begin{minipage}[b]{8.8cm}
\begin{picture}(8.8,7.7)
{\includegraphics[angle=90,viewport=320 430 570 720,clip,width=86mm]{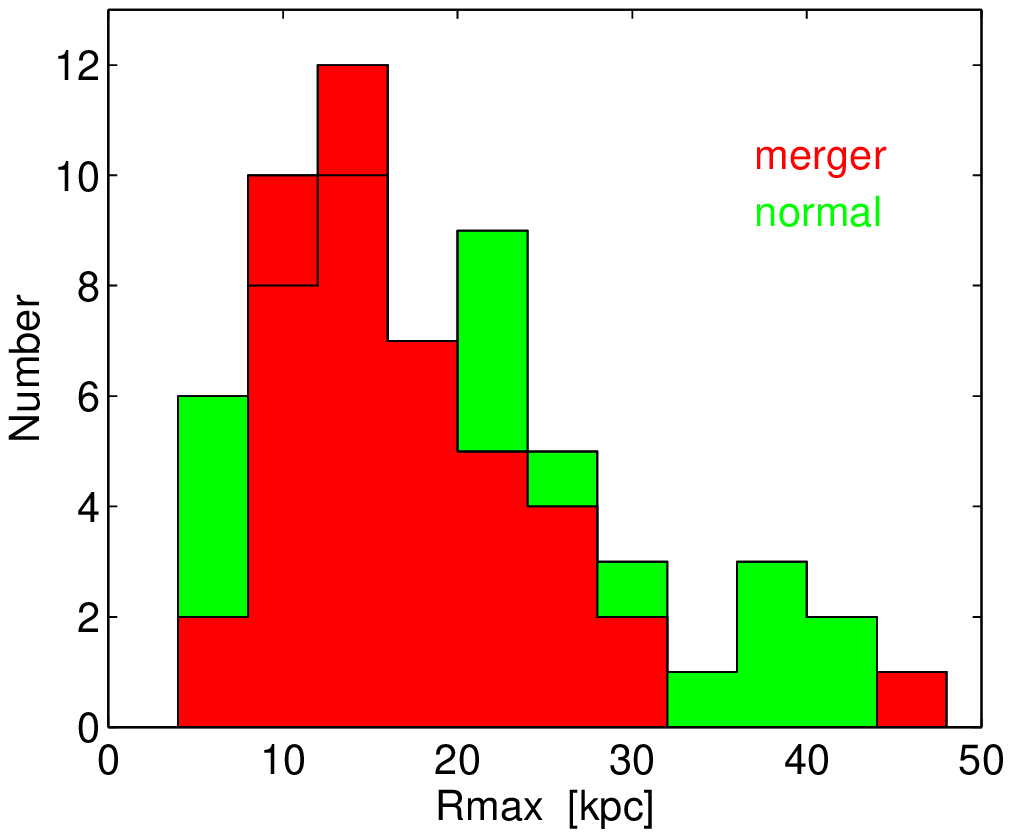}}
\end{picture}

\vspace*{0mm}

{\bf \noindent Fig.~2.} The distribution of disk cut-off radii for the sample of
non-interacting and interacting/merging galaxies.
\end{minipage}
\hfill
\begin{minipage}[b]{8.8cm}
\begin{picture}(8.8,7.7)
{\includegraphics[angle=90,viewport=320 430 570 720,clip,width=86mm]{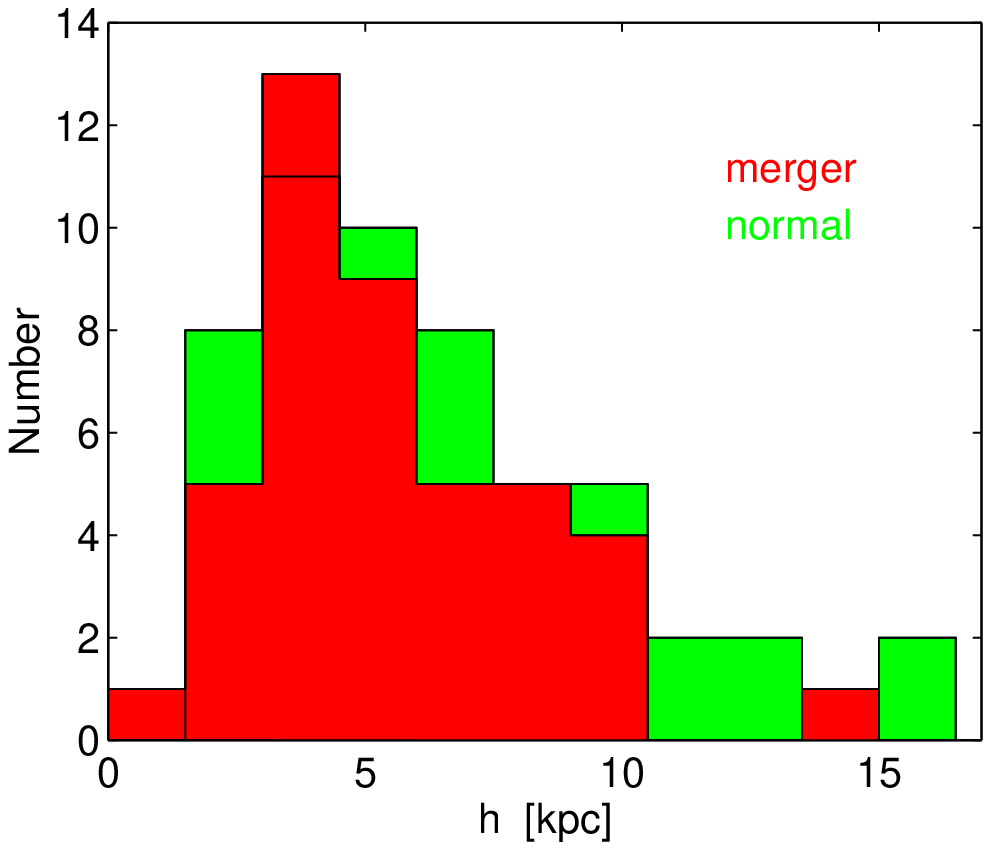}}
\end{picture}

\vspace*{0mm}

{\bf \noindent Fig.~3.} The distribution of disk scale lengths for the sample of
non-interacting and interacting/merging galaxies.
\end{minipage}

\end{figure*}

%%%%%%%%%%%%%%%%%%%%%%%%%%%%%%%%%%%%%%%%%%%%%%%%%%%%%%%%%%%%%%%%%%%

%%%%%%%%%%%%%%%%%%%%%%%%%%%%%%%%%%%%%%%%%%%%%%%%%%%%%%%%%%%%%%%%%%%

\begin{figure*}[t]

\vspace*{74mm}

\begin{minipage}[b]{8.8cm}
\begin{picture}(8.8,7.7)
{\includegraphics[angle=90,viewport=320 430 570 720,clip,width=86mm]{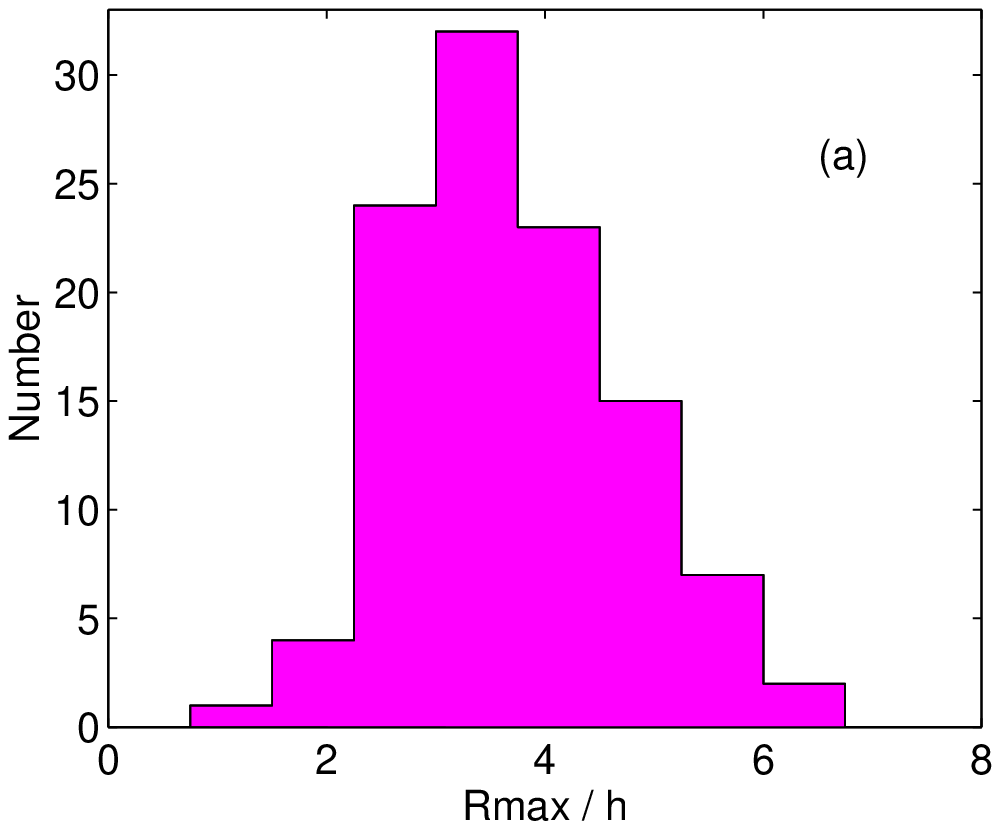}}
\end{picture}
\end{minipage}
\hfill
\begin{minipage}[b]{8.8cm}
\begin{picture}(8.8,7.7)
{\includegraphics[angle=90,viewport=320 430 570 720,clip,width=86mm]{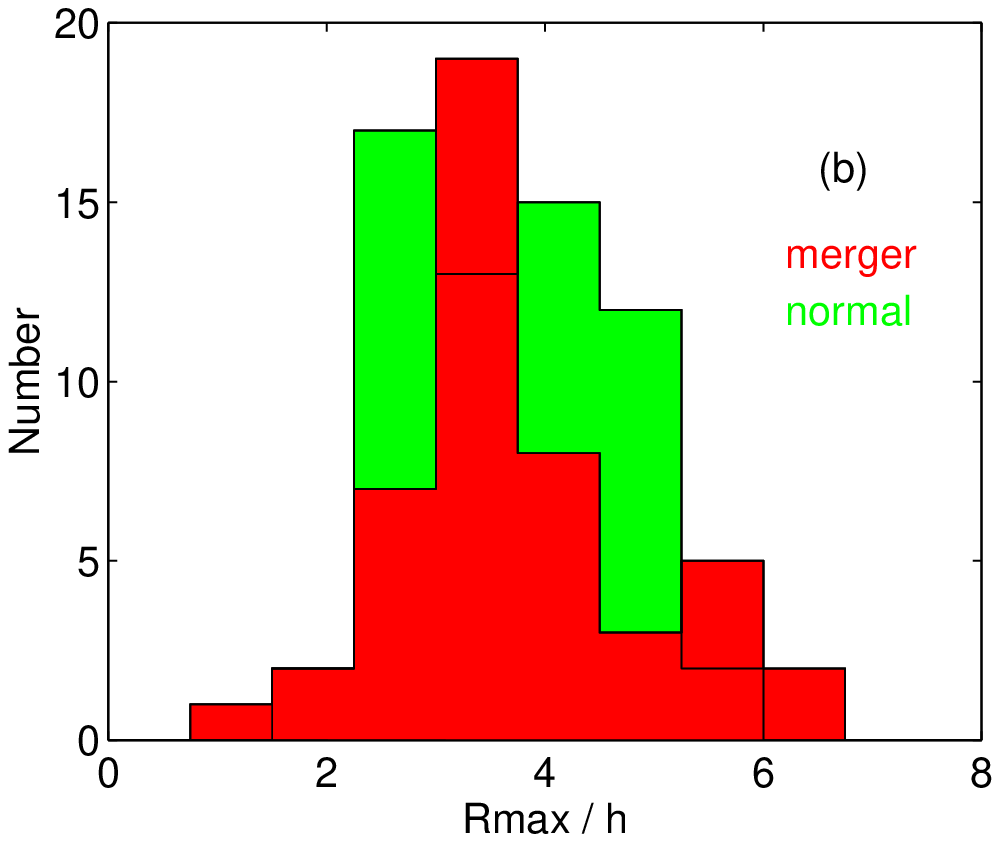}}
\end{picture}
\end{minipage}

\vspace*{-1mm}

{\bf \noindent Fig.~4a and b.} Ratio of radial disk parameters $\Rmax/h$
{\bf(a)} for the total galaxy sample, and {\bf(b)} for normal/interacting galaxies.

\end{figure*}

%%%%%%%%%%%%%%%%%%%%%%%%%%%%%%%%%%%%%%%%%%%%%%%%%%%%%%%%%%%%%%%%%%%

%%%%%%%%%%%%%%%%%%%%%%%%%%%%%%%%%%%%%%%%%%%%%%%%%%%%%%%%%%%%%%%%%%%

%__________________________________________________________________
%
% Table 1 - comparison of radial disk parameter ratios Rmax/h (this study, literature)
%
%__________________________________________________________________
%

\tabcolsep3.9mm

  \begin{table*}[t]
  \caption[ ]
  {Comparison of radial disk parameter ratios $\Rmax/h$ (this study and literature data). \\
   columns: (1) Source: KS~I--III= van der Kruit \& Searle (\cite{kruit1981a},\cite{kruit1981b};
   \cite{kruit1982a}); K= van der Kruit (\cite{kruit1986}); H=Habing (\cite{habing1988});
   SG= Shaw \& Gilmore (\cite{shaw1989}); G= Gilmore et al. (\cite{gilmore1989});
   BT= Barnaby \& Thronson (\cite{barnaby1992}); BD= Barteldrees \& Dettmar (\cite{barteldrees1994});
   SD~I= Schwarzkopf \& Dettmar (\cite{schwarzkopf1997}); SD~II= this study;
   (2) Ratio $\Rmax/h$ and error; (3) Absolute range of $\Rmax$; (4) Mean value of column~(3);
   (5) Number of galaxies used; (6) Range of morph. galaxy types; (7) Distance range of sample galaxies.}
  \label{rmax_h}
  \smallskip
  \begin{flushleft}
  \begin{tabular}{llrccrr}
  \cline{1-7}
  \hline\hline\noalign{\smallskip}
   \multicolumn{1}{c}{Source}       & \multicolumn{1}{c}{$\Rmax/h$} & \multicolumn{1}{c}{$\Rmax$} &
   \multicolumn{1}{c}{$\Rmaxm$}     & \multicolumn{1}{c}{Sample}    & \multicolumn{1}{c}{Type}    &
   \multicolumn{1}{c}{Dist}   \\
    &  &  \multicolumn{1}{c}{[kpc]} & \multicolumn{1}{c}{[kpc]} & & & \multicolumn{1}{c}{[Mpc]}  \\
   \noalign{\smallskip}
   \multicolumn{1}{c}{(1)} & \multicolumn{1}{c}{(2)} & \multicolumn{1}{c}{(3)} &
   \multicolumn{1}{c}{(4)} & \multicolumn{1}{c}{(5)} & \multicolumn{1}{c}{(6)} &
   \multicolumn{1}{c}{(7)} \\
  \noalign{\smallskip}
  \hline\noalign{\smallskip}
  KS~I--III                    & $4.25\pm0.6$ &  7.8 -- 24.9 &17.1&  8 &   3.0 -- 6.0 &  5.0 --
  \hspace{0.7mm} 14.0 \\
  K, H, SG, G                  & $4.20\pm1.0$ &         18.0 & 18.0 & Galaxy   &    4.0 &  --  \\
  BT                           &  3.38        &         19.3 & 19.3 & NGC 5907 &    6.0 & 11.0 \\
  BD                           & $3.70\pm1.0$ & 10.5 -- 38.9 & 20.2 & 27 & 0.0 -- 7.0 & 26.0 -- 113.0 \\
  SD~I                         & $3.61\pm0.9$ & 10.5 -- 37.8 & 19.4 & 37 &-2.0 -- 6.3 & 25.0 -- 162.9 \\
  \noalign{\medskip}
  SD~II (total sample)$\A$     & $3.59\pm0.6$ &  2.9 -- 45.5 &17.2& 108 &    0 -- 9.0 &  3.8 -- 193.0 \\
  SD~II (non-interacting)$\A$  & $3.66\pm0.7$ &  4.2 -- 40.9 &17.2&  61 &    0 -- 9.0 &  3.8 -- 114.4 \\
  SD~II (interacting/merging)$\A$
                               & $3.53\pm0.6$ &  2.9 -- 45.5 &17.0&  47 &    0 -- 9.0 &  4.8 -- 193.0 \\
  \noalign{\smallskip}
  \hline
  \end{tabular}
  \begin{list}{}{}
  \item[$\A$] Optical ($R$-band) data. \\
  \end{list}
  \end{flushleft}
  \end{table*}

%%%%%%%%%%%%%%%%%%%%%%%%%%%%%%%%%%%%%%%%%%%%%%%%%%%%%%%%%%%%%%%%%%%

%__________________________________________________________________

\section{Results}

The following statistical analysis compares the radial and vertical disk parameters of both
galaxy samples based on the optical ($R$-band) data set. The derived main disk parameters are:
inclination angle $i$, cut-off radius $\Rmax$, scale length $h$, scale height $\zo$, and
best-fitting vertical model $f(z)$. They are listed in Table~\ref{parameters}, columns (7)-(11).

%__________________________________________________________________

\subsection{The radial and vertical disk structure}

\subsubsection{The cut-off radius ``\Rmax''}

The distribution of disk cut-off radii derived from the values in Table~\ref{parameters} is shown
in Fig.~2 for the samples of non-interacting and interacting/merging galaxies. Both distributions
cover a wide range between 4~kpc $\le \Rmax \le$ 45~kpc, with the same global maximum around
$\approx 14$~kpc. Slight differences can be detected in a region of large cut-off radii:
many of the interacting/merging galaxy disks are concentrated in a strong peak between
8~kpc $\le \Rmax \le$ 16~kpc, followed by a noticable drop-off towards larger radii.
The distribution ends abruptly at $\Rmax \approx 32$~kpc.

Disks of non-interacting galaxies show a more regular distribution, decreasing from the
maximum at $\Rmax=14$~kpc towards larger radii. However, the median derived for both
distributions is almost identical at $(\Rmax)_{\rm norm.} \approx$ 17.2~kpc and
$(\Rmax)_{\rm merg.} \approx 17.0$~kpc, respectively.

Since the slight differences detected between the two distributions are caused by only a few
galaxies, the KS-test shows that both samples are close to unity and thus statistically
indistinguishable (the result of 0.14 is clearly below the critical 20\%-limit of 0.22).

%Hence, the slight differences between both distributions may indicate a small radial shrinking of
%galactic disks affected by interaction or minor merger by $\approx 10\%$ on average. According
%to the statistical test of Kolmogoroff \& Smirnoff (Sachs \cite{sachs1992}) this radial shrinking
%is, however, below the critical 20\%-limit of the test and thus statistically insignificant.

%__________________________________________________________________

\subsubsection{The disk scale length ``$h$''}

In Fig.~3 the distribution of disk scale lengths is shown for the non-interacting
and interacting/merging galaxy samples. As in the cut-off statistics, both distributions
have approximately the same global maxima, located at $h \approx 4$~kpc.
The disks of non-interacting galaxies possess scale lengths in a wide range
between 1.5~kpc $\le h \le$ 16~kpc, with a regular decrease towards larger values.
The distribution of disk scale lengths of interacting spirals shows a similar behaviour,
but with a more sharply truncated end at $h \approx$ 10.5~kpc.
%The concentration around their maximum at $h \approx 4$ kpc shows analogies to a similar
%feature around the centre of the cut-off statistics mentioned above.
This results in a slightly different median for both distributions at
$h_{\rm norm.} \approx 5.6$~kpc and $h_{\rm merg.} \approx 4.9$~kpc, respectively.

In analogy to the cut-off-statistics, the differences between both $h$-distributions are
only due to a few galaxies (most probably reflecting the slightly different number of galaxies
in both samples) and thus marginal. The KS-test therefore classifies these differences as
statistically insignificant with a result of 0.11, which is clearly below the mentioned
20\%-limit of 0.22.

%__________________________________________________________________

\subsubsection{The ratio of radial parameters ``$\Rmax/h$''}

Although several studies of edge-on spiral galaxies were focussed on an investigation of disk
scale parameters of non-interacting galaxies, only a few have studied the ratio of disk
cut-off radius to the scale length $\Rmax/h$ (KS I-III; Barnaby \& Thronson \cite{barnaby1992};
Barteldrees \& Dettmar \cite{barteldrees1994}; Schwarzkopf \cite{schwarzkopf1999}; Schwarzkopf
\& Dettmar \cite{schwarzkopf1997}).
One of the reasons might be the fact that for most purposes a disk model consisting of a
radial exponentially decreasing luminosity without a cut-off may be a good approximation
(Shaw \& Gilmore \cite{shaw1989}; de Grijs \& van der Kruit \cite{grijs1996}; de Grijs et al.
\cite{grijsetal1997}; Reshetnikow \& Combes \cite{reshetnikov1996}, \cite{reshetnikov1997}).
However, the importance of a cut-off radius as a reasonable step towards a more realistic
description of the properties of galactic disks and its necessity for a precise quantitative
description of the observed radial disk profiles were impressively confirmed by the results
of the former studies. The existence of a disk cut-off, although still objectionable within
the framework of galaxy evolution, seems therefore now well accepted.

This is consistent with the results of our disk modelling procedure (Paper~I), showing that
the shape of radial disk profiles and thus the derived value for the disk scale length is also
influenced by the size of the cut-off radius. Using a disk model without a cut-off would
increase the (existing) uncertainties in estimating reliable disk scale lengths.

The ratios $\Rmax/h$ found in this study for both samples of non-interacting and
interacting/merging galaxies are summarized in Table~\ref{rmax_h}. They are compared with
the data from literature of published samples of non-interacting galaxies. Fig.~4 shows the
$\Rmax/h$-distribution for both the total galaxy sample and the two subsamples.
Due to its statistically significance (108 galaxies) the distribution of the total galaxy
sample is very regular and, though the slightly steeper drop-off towards smaller
values, close to a Gaussian. The median of the total sample is at $(\Rmax/h)_{\rm tot} = 3.59$,
that of the non-interacting and interacting/merging galaxy sample at $(\Rmax/h)_{\rm norm.} = 3.66$
and  $(\Rmax/h)_{\rm merg.} = 3.53$, respectively (typical errors are given in Table~\ref{rmax_h}).
According to the KS-test both distributions are statistically indistinguishable with a result of
0.12 (critical 20\%-limit is at 0.21).

The relatively high scatter of the merger-sample, which becomes apparent in a wider basis of the
distribution (Fig.~4b), is mainly due to a radially perturbed disk structure of these galaxies.
These perturbations may cause asymmetries in the radial disk profiles and thus substantially errors
in the scale length and/or the cut-off radius. Unlike this result it is a remarkable fact that the
$\Rmax/h$-distribution of interacting/merging galaxies shows a very sharp peak at
$\Rmax/h \approx 3.4$, while non-interacting galaxies are spread in a much wider plateau.

Within the estimated errors the obtained $\Rmax/h$-ratios of all (sub-)samples are consistent
with other studies (Table~\ref{rmax_h}). However, this study -- dealing with the by far largest
galaxy sample for a cut-off statistics and hence with smaller errors -- indicates that the
$\Rmax/h$-ratios are lower than previously argued. The mean ratio now seems to be closer to
$(\Rmax/h)_{\rm tot} = 3.6$ than to the value 4.2 often used in the literature.
This fact may be explained by selection effects of studies dealing with statistically very small
samples of relatively nearby ($D \le 20$ Mpc) and medium-sized ($\Rmax \le 25$ kpc) galaxies, which
is the case for some of these data. Therefore, some additional information on these studies are
included in Table~\ref{rmax_h}. A possible correlation between disk parameters and distance will
be studied in Sect.~3.2.

%%%%%%%%%%%%%%%%%%%%%%%%%%%%%%%%%%%%%%%%%%%%%%%%%%%%%%%%%%%%%%%%%%%

\begin{figure*}[t]

\vspace*{74mm}

\begin{minipage}[b]{8.8cm}
\begin{picture}(8.8,7.7)
{\includegraphics[angle=90,viewport=320 430 570 720,clip,width=86mm]{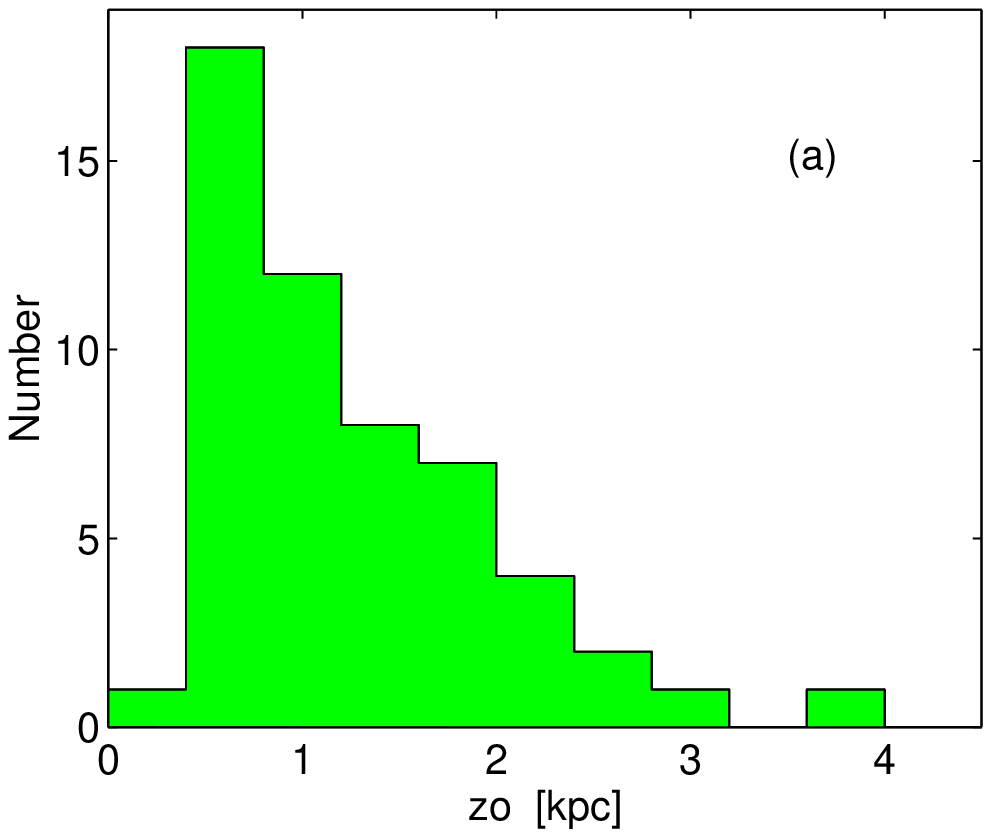}}
\end{picture}
\end{minipage}
\hfill
\begin{minipage}[b]{8.8cm}
\begin{picture}(8.8,7.7)
{\includegraphics[angle=90,viewport=320 430 570 720,clip,width=86mm]{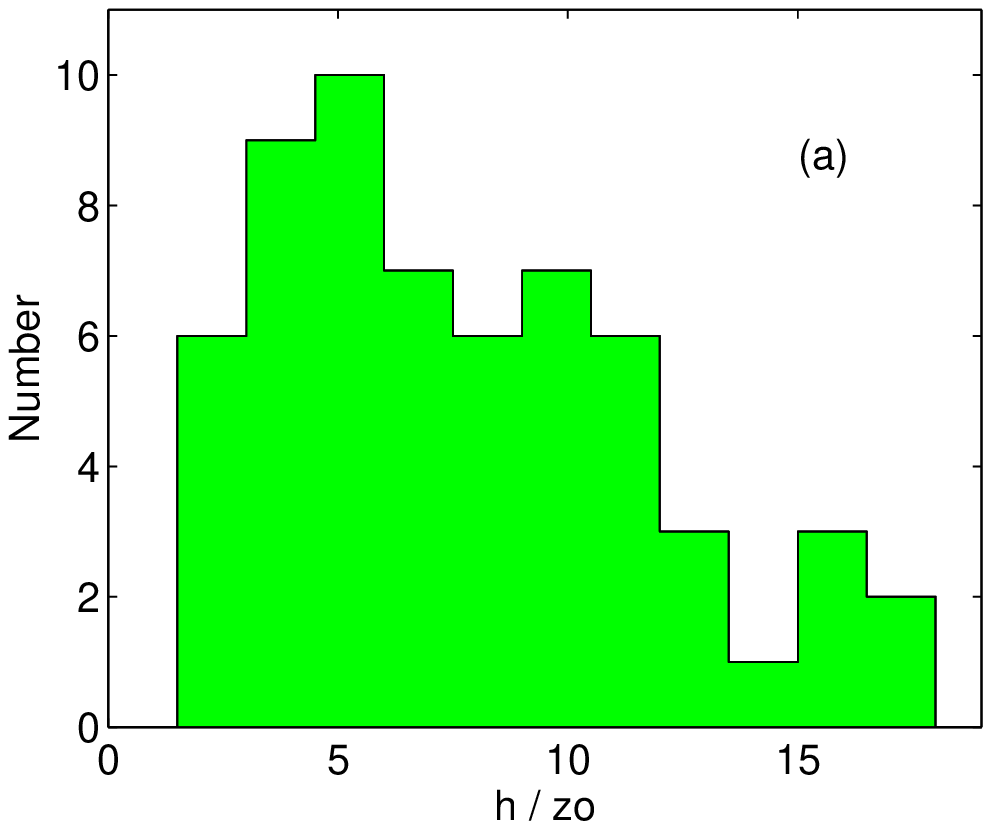}}
\end{picture}
\end{minipage}

%%%%%%%%%%%%%%%%%%%%%%%%%%%%%%%%%%%%%%%%%%%%%%%%%%%%%%%%%%%%%%%%%%%

\vspace*{74mm}

%%%%%%%%%%%%%%%%%%%%%%%%%%%%%%%%%%%%%%%%%%%%%%%%%%%%%%%%%%%%%%%%%%%

\begin{minipage}[b]{8.8cm}
\begin{picture}(8.8,7.7)
{\includegraphics[angle=90,viewport=320 430 570 720,clip,width=86mm]{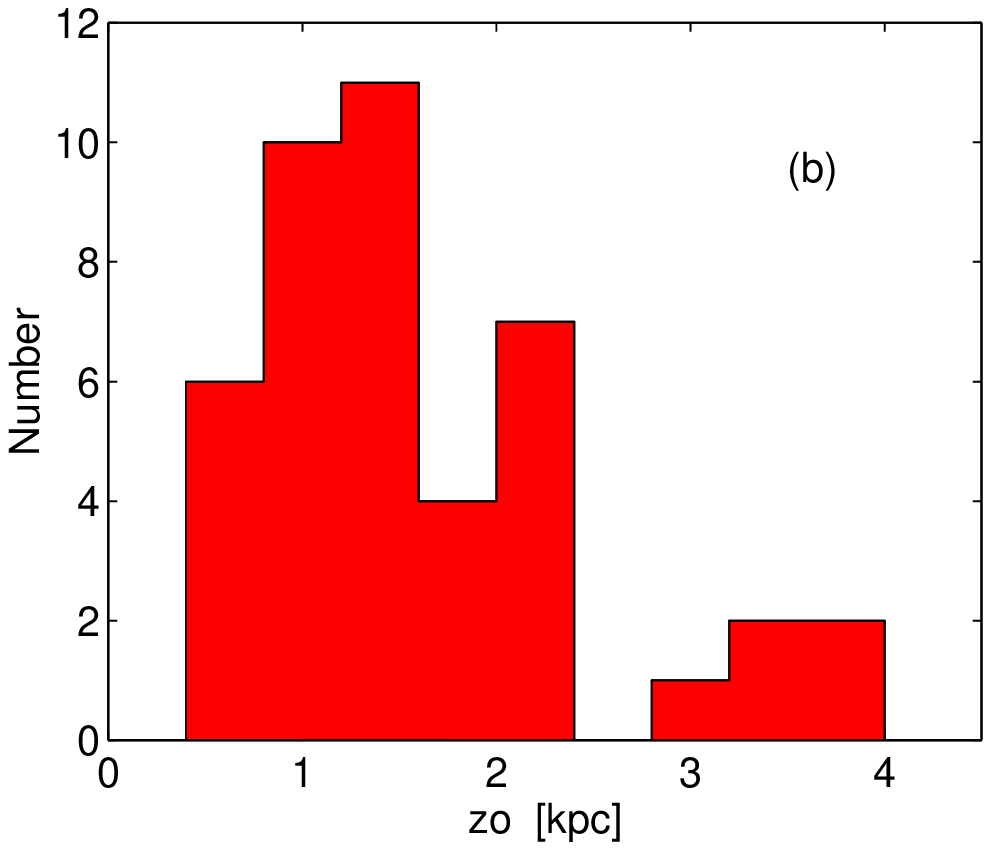}}
\end{picture}

\vspace*{0mm}

{\bf \noindent Fig.~5a and b.} The distribution of disk scale heights {\bf(a)} for the
sample of non-interacting and {\bf(b)} for interacting/merging galaxies.
\end{minipage}
\hfill
\begin{minipage}[b]{8.8cm}
\begin{picture}(8.8,7.7)
{\includegraphics[angle=90,viewport=320 430 565 720,clip,width=86mm]{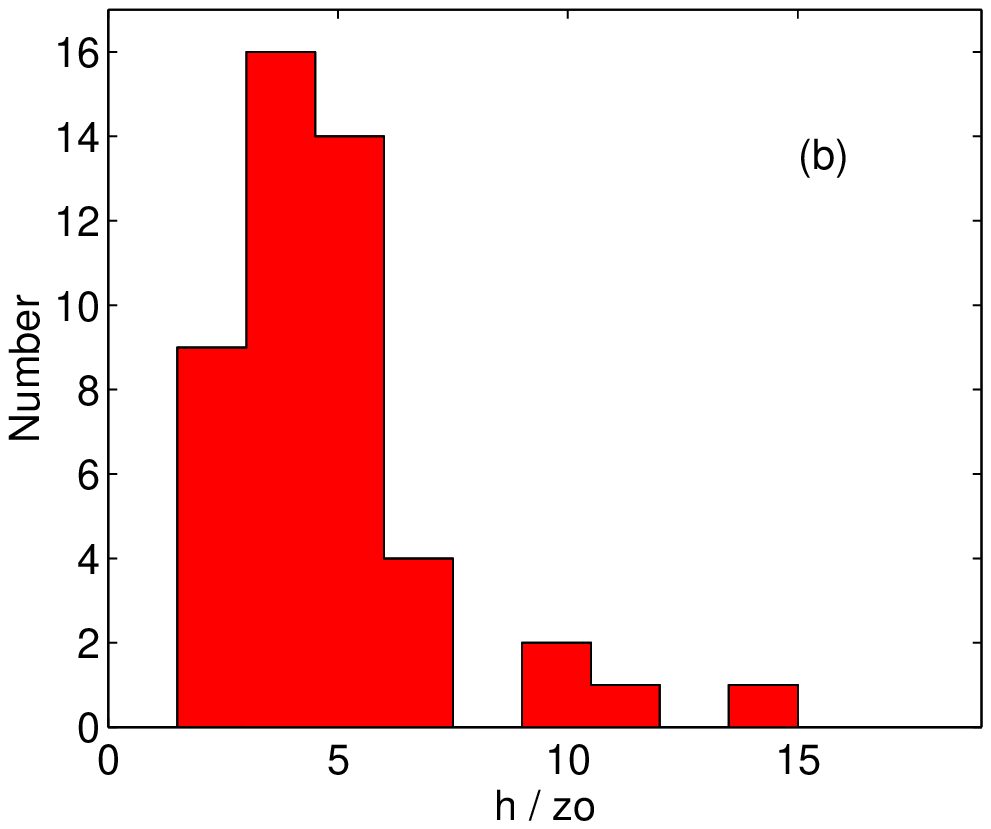}}
\end{picture}

\vspace*{0mm}

{\bf \noindent Fig.~6a and b.} The ratio of radial to vertical disk parameters $h/\zo$
{\bf(a)} for the sample of non-interacting and {\bf(b)} for interacting/merging galaxies.
\end{minipage}

\end{figure*}

%%%%%%%%%%%%%%%%%%%%%%%%%%%%%%%%%%%%%%%%%%%%%%%%%%%%%%%%%%%%%%%%%%%

%__________________________________________________________________

%%%%%%%%%%%%%%%%%%%%%%%%%%%%%%%%%%%%%%%%%%%%%%%%%%%%%%%%%%%%%%%%%%%

\begin{figure*}[t]

\vspace*{78mm}

\begin{minipage}[b]{8.8cm}
\begin{picture}(8.8,8.2)
{\includegraphics[angle=90,viewport=0 50 610 725,clip,width=91.5mm]{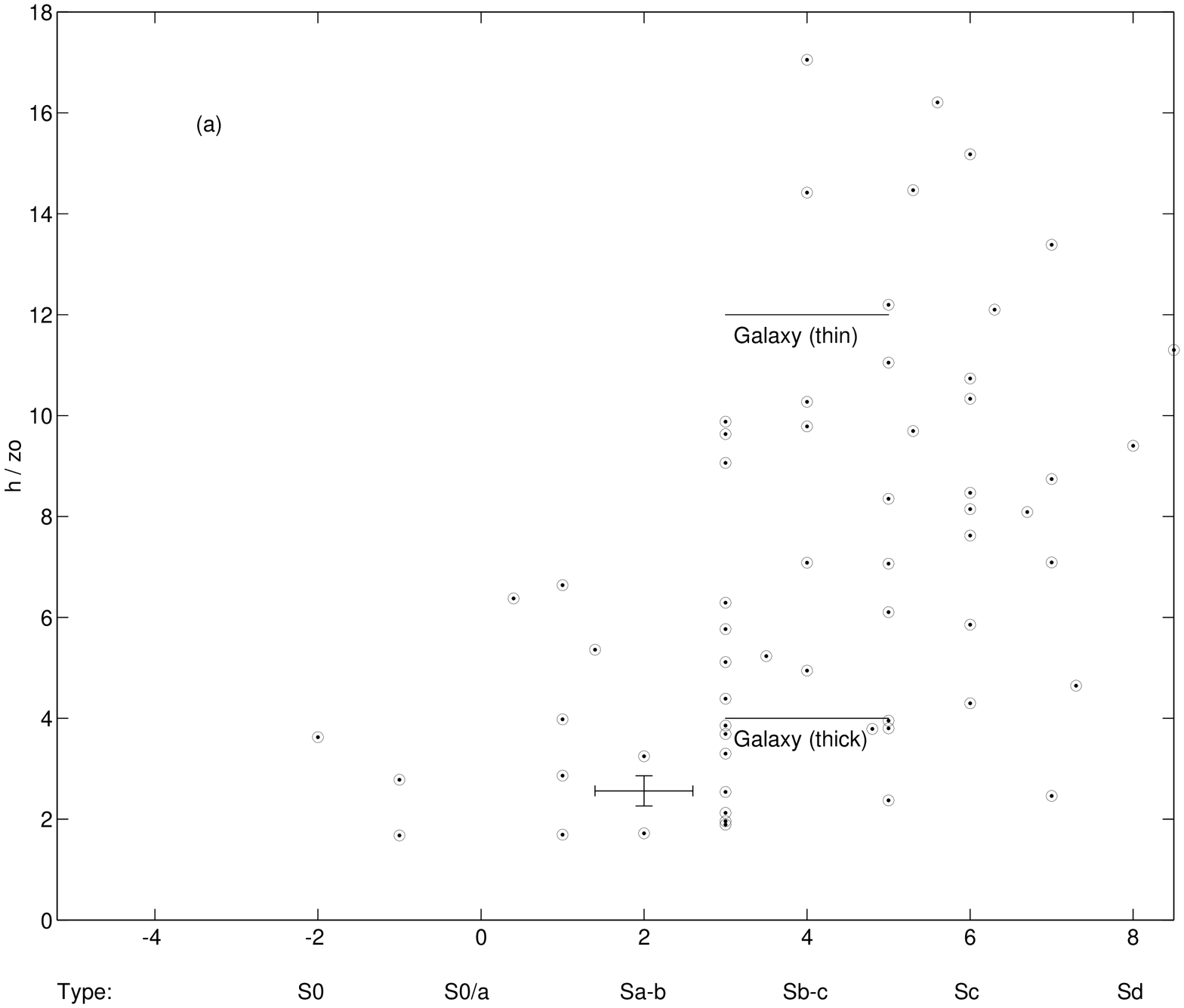}}
\end{picture}
\end{minipage}
\hfill
\begin{minipage}[b]{8.8cm}
\begin{picture}(8.8,8.2)
{\includegraphics[angle=90,viewport=0 50 610 725,clip,width=91.5mm]{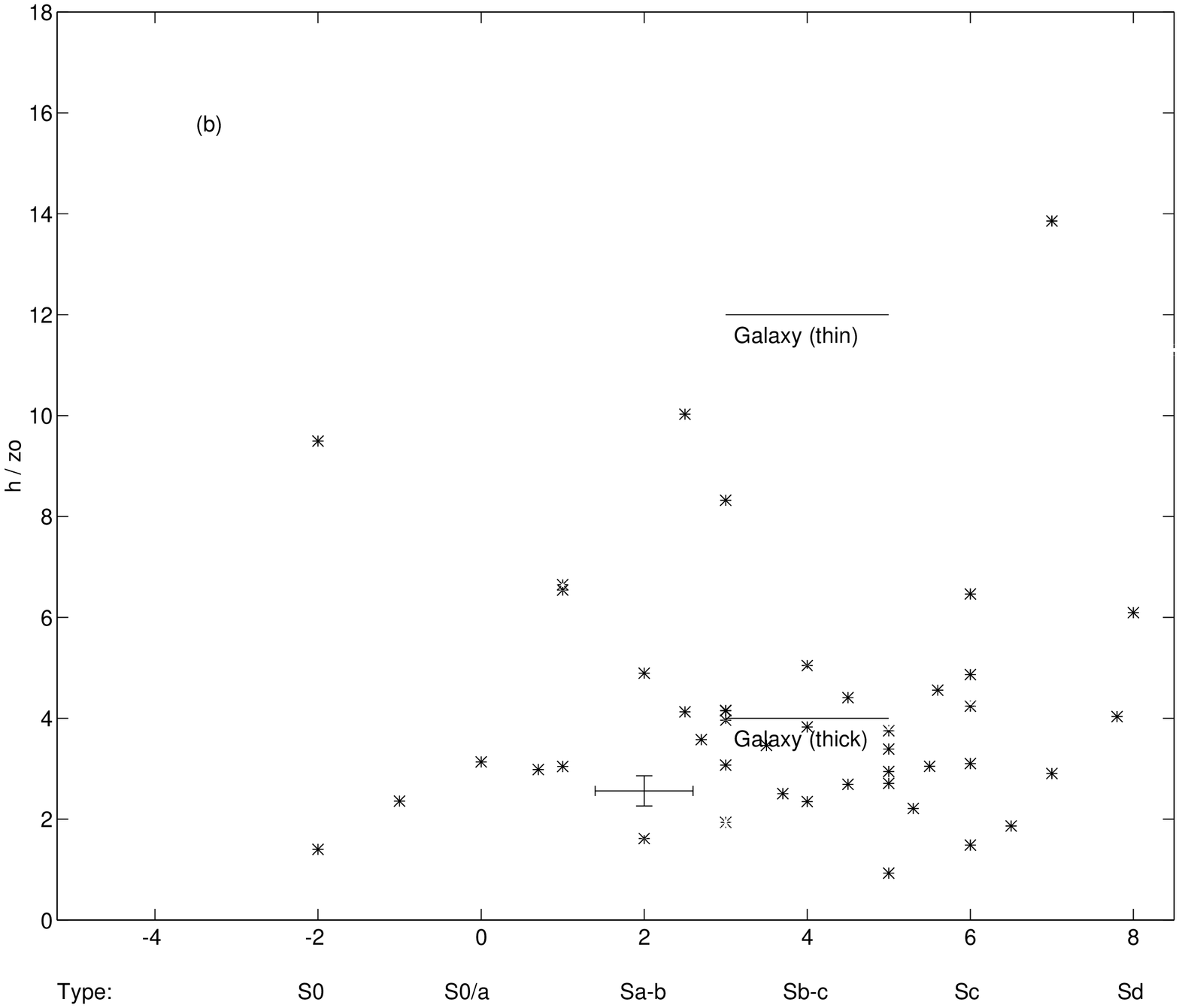}}
\end{picture}
\end{minipage}

\vspace*{0mm}

{\bf \noindent Fig.~7a and b.} The dependence of ratio $h/\zo$ on galaxy type
{\bf(a)} for the sample of non-interacting galaxies. The position of the thin/thick
disk component of our Galaxy as well as typical errors are indicated. Disks of
interacting/merging galaxies {\bf(b)} show a significant lower ratio and
do not follow the $h/\zo$-trend of non-interacting galaxies.

\end{figure*}

%%%%%%%%%%%%%%%%%%%%%%%%%%%%%%%%%%%%%%%%%%%%%%%%%%%%%%%%%%%%%%%%%%%

\subsubsection{The disk scale height ``$\zo$''}

The distribution of exponential disk scale heights $\zo$ -- calculated by using the absolute
$\zo$-values in Table~\ref{parameters}, column (10), and the corresponding transformations in
(5) of Paper I -- is shown in Fig.~5a for the non-interacting galaxy sample, and in Fig.~5b for
the interacting/merging galaxies. By comparing these diagrams clear differences concerning both
the trend and the median of the distributions can be detected:

The majority ($\approx 60\%$) of normal galaxies is concentrated in a region $\zo \le 1.1$ kpc,
with a clear maximum between 400 pc $\le \zo \le$ 800 pc. The distribution shows a sharp truncation
towards extremely thin disks at $\zo = 300$ pc, followed by a very regularly decreasing part
towards thicker disks up to $\zo \approx 3$ kpc.

Unlike this, the distribution of interacting/merging galaxies shows a trend that
is completely contrary to that found for normal galaxies: the scale height increases
rapidly towards thicker disks, having a maximum between 1.2 kpc $\le \zo \le$ 1.6 kpc,
and a higher frequency of disks thicker than 2 kpc. The median for both distributions
is $(\zo)_{\rm norm.} \approx 1.0$ kpc and $(\zo)_{\rm merg.} \approx 1.5$ kpc.
It is also remarkable that very thin disks in the range 250~pc $\le \zo \le$ 450~pc, as
they are frequently observed in the sample of non-interacting galaxies (e.g. UGC 231,
UGC 4278, UGC 4943, see Table~\ref{parameters}), are completely missing in the interacting
sample. Since this is not due to the sample selection criteria (Paper~I) it is most likely
a result of the various collective instabilities of such thin disks, in particular against
external perturbations as they are commonly evoked by tidal interactions or even minor
mergers (Gerin et al. \cite{gerin1990}; Toth \& Ostriker \cite{toth1992}).

The KS-test confirms that the differences between both distributions are statistically
significant: the result of 0.24 is above both the 20\% (0.22) and the 15\% (0.23) limit of the
test.
Since it was shown (Fig.~1 of Paper~I and Sect.~2 of this paper, resp.) that the absolute disk
parameters of both galaxy samples are not affected by selection biases -- the distance- and type
distribution were found almost indistinguishable -- it can be concluded that galactic disks
affected by interactions/minor mergers are $\approx 1.5$ times thicker on average. We therefore
infer that the disk thickening is in fact caused by the interaction or the merging process.

However, it is not yet clear whether this thickening effect was evoked only by a locally
increased scale height due to a vertically perturbed disk structure or by global disk
thickening. This can be clarified only after a detailed analysis of the vertical disk structure.
Therefore, the behaviour of the disk scale height with radial distance, i.e. $\zo(R)$, and the
effects of vertical disk perturbations will be investigated in detail in a forthcoming paper
(Paper~III, Schwarzkopf \& Dettmar \cite{schwarzkopf2000c}).

%%%%%%%%%%%%%%%%%%%%%%%%%%%%%%%%%%%%%%%%%%%%%%%%%%%%%%%%%%%%%%%%%%%

%__________________________________________________________________
%
% Table 2 - The dependence of disk parameters on distance.
%
%__________________________________________________________________
%

\tabcolsep3.4mm

  \begin{table*}[t]
  \parbox{16.5cm}{
  \caption[ ]{The dependence of measured disk parameters on distance and morphological galaxy type. \\
   columns: (1) Disk parameter used for the statistics (given as median of the distance-subsample
   and in per cent of the median of the total sample (last column)); (2) Distance range (in Mpc),
   number of objects ($n$) within this range, and corresponding morph. galaxy type $T$ (given as
   median for each subsample and the total sample, resp.).}}
  \label{distance}
  \begin{flushleft}
  \begin{tabular}{llcrrcrrcrrcc}
  \cline{1-13}
  \hline\hline\noalign{\smallskip}
  \multicolumn{2}{c}{Disk}            &&  \multicolumn{10}{c}{Dist [Mpc]}  \\
  \noalign{\smallskip}
  \cline{4-13}
  \noalign{\smallskip}
  \multicolumn{2}{c}{parameter$\A$}   &&  \multicolumn{2}{c}{0 -- 22}     &&
  \multicolumn{2}{c}{22 -- 45.5}      &&  \multicolumn{2}{c}{45.5 -- 193} &&
  \multicolumn{1}{c}{3.8 -- 193}  \\
  \noalign{\smallskip}
  \multicolumn{2}{c}{}                &&  \multicolumn{2}{c}{($n=32$)}    &&
  \multicolumn{2}{c}{($n=32$)}        &&  \multicolumn{2}{c}{($n=33$)}    &&
  \multicolumn{1}{c}{(total sample)}  \\
%  \noalign{\smallskip}
%  \multicolumn{2}{c}{}                  &&  \multicolumn{2}{c}{($t \approx 4.8$)}    &&
%  \multicolumn{2}{c}{($T \approx 5.1$)} &&  \multicolumn{2}{c}{($T \approx 3.4$)}    &&
%  \multicolumn{1}{c}{($T \approx 3.9$)}  \\
  \noalign{\smallskip}
  \multicolumn{2}{c}{(1)}             &&  \multicolumn{10}{c}{(2)} \\
  \noalign{\smallskip}
  \hline\noalign{\smallskip}
  \multicolumn{2}{c}{}                   &&  \multicolumn{2}{c}{$T = 4.8 \pm 0.9$}  &&
  \multicolumn{2}{c}{$T = 5.1 \pm 0.9$}  &&  \multicolumn{2}{c}{$T = 3.4 \pm 0.9$}  &&
  \multicolumn{1}{c}{$T = 3.9 \pm 0.7$}  \\
  \noalign{\smallskip}
  \hline\noalign{\medskip}
   $\Rmax$ & [kpc]              && 11.00 & (64\%) && 14.90 &  (87\%) && 23.50 & (137\%) && $17.20 \pm 2.6$
   \hspace{0.9mm} \\
   $h$     & [kpc]              &&  3.58 & (67\%) &&  4.94 &  (93\%) &&  7.75 & (145\%) && $ 5.33 \pm 0.8$ \\
   $\zo$   & [kpc]              &&  0.98 & (80\%) &&  1.10 &  (89\%) &&  1.87 & (152\%) && $ 1.23 \pm 0.2$ \\
  \noalign{\medskip}
  \multicolumn{2}{l}{$\Rmax/h$} &&  3.50 & (97\%) &&  3.79 & (106\%) &&  3.43 &  (96\%) && $ 3.59 \pm 0.4$ \\
  \multicolumn{2}{l}{$h/\zo$}   &&  5.00 & (93\%) &&  5.25 &  (97\%) &&  5.67 & (105\%) && $ 5.40 \pm 0.9$ \\
  \noalign{\smallskip}
  \hline\noalign{\smallskip}
  \end{tabular}
  \begin{list}{}{}
  \item[$\A$] Optical ($R$-band) data. \\
  \end{list}
  \end{flushleft}
  \end{table*}

%%%%%%%%%%%%%%%%%%%%%%%%%%%%%%%%%%%%%%%%%%%%%%%%%%%%%%%%%%%%%%%%%%%

%__________________________________________________________________

\subsubsection{The ratio of radial to vertical parameters ``$h/\zo$''}

The ratio of radial to vertical disk scale parameters $h/\zo$ -- hence a normalized
thickness -- is very suitable for characterizing and comparing the disk structure of
edge-on galaxies independently of their distance. It is thus -- unlike the absolute
values of scale heights -- more reliably for a detection of small changes in the
vertical disk structure.

Fig.~6 shows the $h/\zo$-distribution for both samples of non-interacting and
interacting/merging galaxies. As it can be seen clearly, the ratio $h/\zo$ for
normal galaxies covers a wide range $2 \le h/\zo \le 18$ between extreme thick
and thin disks. Most of these galaxies are concentrated between $2 \le h/\zo \le 12$,
with a maximum at $h/\zo \approx 5.3$. There is only a slight decrease in the number
of flat disks from the maximum of the distribution towards extreme flat disk ratios,
i.e. between $6 \le h/\zo \le 12$.

Unlike this, the most striking feature in the $h/\zo$-distribution for interacting/merging
galaxies is a very sharp concentration between $2 \le h/\zo \le 6$, while the flat
disks with ratios typically $h/\zo > 7$ are completely missing. The distribution peaks at
$h/\zo \approx 3.7$, and there is no smooth transition towards thicker or thinner disks
on both sides of this sharply truncated distribution.

The ratio of the median values for both distributions and hence a lower limit for vertical disk
thickening is $(h/\zo)_{\rm norm:merg} =  7.1:4.3 \approx 1.7$. This factor, however, considerably
underestimates the differences between both distributions due to the above mentioned lack of
(extremely) thin disks ratios. The differences between both distributions are, according to the
KS-test, statistically significant with a result of 0.41 even if the strongest test criterium,
the 0.1\%-level (limit 0.38), is used.

Together with the nearly unchanged scale lengths (Fig.~3) and the differences found between
the absolute values of disk scale heights (Fig.~5) it can be concluded that vertical thickening
of galactic disks affected by interactions/minor mergers amounts to $\approx 70\%$.
The changes of the disk structure result mainly from an increase in scale height.

Finally, Fig.~7a shows that the ratio $h/\zo$ of non-interacting galaxies correlates with the
morphological type of galaxies, in the sense that the disks become systematically thinner from
early types (S0, $h/\zo \approx 2 \pm 2$) to late types (Sc/Sd, $h/\zo \approx 8 \pm 2$). Despite
the relatively high intrinsic scatter, there is a smooth transition between these two extremes.
It should be stressed that -- due to the corrections necessary in order to compare disk scale
heights derived from different vertical luminosity distributions (exp, sech, and $\iso$) --
the ratio $h/\zo$ can be higher than 10 for some of the disks. This value represents the maximum
theoretical value allowed for stable disks, derived from the so-called ``maximum disk'' fits
(Bottema \cite{bottema1993}). The results obtained are therefore not unexpected, and comparable
ratios were also found in earlier studies (de Grijs \cite{grijs1997}; de Grijs \& van der Kruit
\cite{grijs1996}; Schwarzkopf \& Dettmar \cite{schwarzkopf1997}; Shaw \& Gilmore \cite{shaw1989}).

In contrast to this there is no such correlation between galaxy type and ratio $h/\zo$ for the
sample of interacting/merging galaxies (Fig.~7b). The latter typically possess thickened and
disturbed disks, and are therefore concentrated in the lower right part of the panel.

For comparison, the position of our Galaxy disk is also indicated in Fig.~7. The disk of the Milky Way
consists presumably of both a thin and a thick component with $h/\zo \approx 12$ and $h/\zo \approx 4$,
respectively (according to Gilmore \& Reid \cite{gilmore1983}).

%__________________________________________________________________

\subsection{The dependence of disk parameters on distance}

As briefly mentioned in Sect.~3.1.3 (Table~\ref{rmax_h}), the samples of some previous
studies seem to be biased towards nearby objects. In order to check if the distance as a free
parameter has any obvious effect on the derived disk parameters of this study, we analyzed 3
sub-samples (no differences were made between interacting and non-interacting galaxies) defined
by various distance ranges (Table~2).

To ensure statistically large enough subsamples the total sample (97 galaxies) was split into three
subsamples containing about 32 galaxies each. According to the non-uniform distribution of distances
(Fig.~1) this leads to different distance intervals. Afterwards, the medians for all disk parameters
in each distance-subsample were estimated. For a better comparison, these values are also given in
per cent of the median of the total sample. Additionally, the averaged morphological galaxy type
is listed for each subsample.

As can be seen in Table~2, there is a clear trend for all absolute disk parameters ($\Rmax, h, \zo$)
in the sense that their median values are increasing with distance. This behaviour reflects, however,
undoubtly a selection effect and is therefore not unusual for all those studies that are using
similar selection criteria: due to the limited spatially resolution of the images the selection of
suitable, remote galaxies is biased towards large and bright objects, disfavouring the relative number
of physically small galaxies. This is confirmed by the strikingly simultaneous trend for all absolute
disk parameters listed, which goes, on average, from 70\% and 90\% to 145\% of the median of the total
sample (Table~2, values in parenthesis). In spite of this trend the distance-independent ratios
$\Rmax/h$ and $h/\zo$ both stay nearly constant over the whole range.

Hence, within the given errors there is no correlation between disk parameters of different distance
intervals and corresponding galaxy type. Since the distance distribution of interacting and
non-interacting galaxies is statistically indistinguishable (Fig.~1b), the observed selection
effect applies to both samples and does therefore not influence the comparison of disk parameters
derived in this study.

%__________________________________________________________________

%%%%%%%%%%%%%%%%%%%%%%%%%%%%%%%%%%%%%%%%%%%%%%%%%%%%%%%%%%%%%%%%%%%

%__________________________________________________________________
%
% Table 3 - Vertical surface brightness distribution of galactic disks.
%
%__________________________________________________________________
%

\tabcolsep1.8mm

  \begin{table}[t]
  \caption[ ]
   {Vertical surface brightness distribution of galactic disks.
   columns: (1) Sample used for the statistics: total= all galaxies; non-int.= non-interacting;
   int./merg.= interacting/merging galaxies; (2) -- (4) Number and percentage of disks
   with vertical luminosity distribution $\propto$ exp,- sech,- and $\iso$; (5) Number
   of galaxies in the (sub-) sample.}
  \label{profiles}
  \medskip
  \begin{flushleft}
  \begin{tabular}{lrrcrrcccc}
  \cline{1-10}
  \hline\hline\noalign{\smallskip}
   \multicolumn{1}{c}{Sample} & \multicolumn{8}{c}{Vertical Luminosity Distribution ...} &
   \multicolumn{1}{c}{Number} \\
   \noalign{\smallskip}
   \multicolumn{1}{c}{}                & \multicolumn{2}{c}{$\propto$ exp}  &&
   \multicolumn{2}{c}{$\propto$ sech} && \multicolumn{2}{c}{$\propto \iso$} &
   \multicolumn{1}{c}{of} \\
   \noalign{\smallskip}
   \cline{2-3}
   \cline{5-6}
   \cline{8-9}
   \noalign{\smallskip}
   \multicolumn{1}{c}{}       & \multicolumn{1}{c}{[n]}  &  \multicolumn{1}{c}{[\%]} &&
   \multicolumn{1}{c}{[n]}    & \multicolumn{1}{c}{[\%]} && \multicolumn{1}{c}{[n]}  &
   \multicolumn{1}{c}{[\%]}   & \multicolumn{1}{c}{Galaxies} \\
   \noalign{\smallskip}
   \multicolumn{1}{c}{(1)}    & \multicolumn{2}{c}{(2)}  && \multicolumn{2}{c}{(3)} &&
   \multicolumn{2}{c}{(4)}    & \multicolumn{1}{c}{(5)}  \\
  \noalign{\smallskip}
  \hline\noalign{\smallskip}
  \multicolumn{10}{c}{\bf Optical Data} \\
  \noalign{\smallskip}
  \hline\noalign{\smallskip}
   total       &  40  &  36 &&  62  &  56  &&  8  &  7  & 110  \\
   non-int.    &  22  &  36 &&  34  &  56  &&  5  &  8  &  61  \\
   int./merg.  &  18  &  37 &&  28  &  57  &&  3  &  6  &  49  \\
  \noalign{\smallskip}
  \hline\noalign{\smallskip}
  \multicolumn{10}{c}{\bf Near Infrared Data} \\
  \noalign{\smallskip}
  \hline\noalign{\smallskip}
   total       &  11  &  27 &&  27  &  66  &&  3  &  7  &  41  \\
   non-int.    &   7  &  35 &&  11  &  55  &&  2  & 10  &  20  \\
   int./merg.  &   4  &  19 &&  16  &  76  &&  1  &  5  &  21  \\
  \noalign{\smallskip}
  \hline
  \end{tabular}
  \end{flushleft}
  \end{table}

%%%%%%%%%%%%%%%%%%%%%%%%%%%%%%%%%%%%%%%%%%%%%%%%%%%%%%%%%%%%%%%%%%%

\subsection{The vertical surface brightness distribution}

The disk models applied in this study use a set of 3 different functions $f(z)$ in order to
describe the luminosity distribution $L(z)$ vertically to the disk plane: $f(z) \propto$ exp, sech,
and $\iso$. These functions were proposed in some fundamental papers by van der Kruit \& Searle
(KS~I-III), Wainscoat et al. (\cite{wainscoat1989}, \cite{wainscoat1990}), and Burkert \& Yoshii
(\cite{burkert1996}). A detailed description of their properties as well as a comparison
between different distributions were given in Sect.~4 of Paper I.

The quantitative results and experiences made in this study after fitting the disk profiles
of about 150 highly-inclined/edge-on galaxies in optical and in near infrared (NIR)
passbands can be summarized as follows (the complete statistics is listed in Table~\ref{profiles}):

\begin{itemize}

\item A combination of 3 different luminosity distributions $f(z)$ allows a very flexible
description of vertical disk profiles of all galaxies observed.

\smallskip

\item The fit quality achieved is better than $\pm \, 0.^{\rm m}2$ for nearly all of the
non-interacting galaxies investigated both in the optical and in NIR, even at small $z$.

\smallskip

\item The fit quality of galaxy disks affected by interaction/minor merger is, in principle,
comparable to that found for non-interacting galaxies. However, some galaxies in the first
sample possess vertical profiles with larger deviations from an ideal disk.

\smallskip

\item These deviations are mainly due to tidal perturbations on short scales and/or a
warped disk. Such features seem to be characteristic for disks in an intermediate stage of
interactions/minor mergers. A detailed study of the vertical disk structure will be
given in Paper~III.

\smallskip

\item The statistics for the best-fitting vertical luminosity distribution $f(z)$ -- applied
to both optical and NIR disk profiles -- is as follows (optical : NIR, Table~\ref{profiles},
columns 2-4): \newline
(56 : 66)\% $\propto$ sech; (36 : 27)\% $\propto$ exp; (7 : 7)\% $\propto \iso$.

\smallskip

\item Thus, the vertical luminosity profiles of nearly all ($\approx 93\%$) of the galaxies
are non-isothermal. In fact the profiles are more sharply peaked and preferentially somewhat
closer to a sech- than to an exp-distribution.

\smallskip

\item Statistically, the fraction of galaxies with a certain vertical luminosity profile
(exp, sech, $\iso$) is independent of the passband investigated (differences $<10\%$).

\smallskip

\item In the optical there is no fundamental difference between vertical disk profiles
of non-interacting and interacting/merging galaxies.

\smallskip

\item Accordingly, almost the same percentage of (non-interacting : interacting) galaxies
shows identical vertical distributions in the optical (Table~\ref{profiles}): \newline
(56 : 57)\% $\propto$ sech; (36 : 37)\% $\propto$ exp; (8 : 6)\% $\propto \iso$.

\smallskip

\item In the NIR, interacting galaxies display preferentially sech-profiles
($76\% \propto$ sech; $19\% \propto$ exp), while in the optical this distribution
is shifted towards the exp-profile ($57\% \propto$ sech; $37\% \propto$ exp).

\smallskip

\item The results obtained are independent of the morphological type of galaxies.

\end{itemize}

%__________________________________________________________________

%%%%%%%%%%%%%%%%%%%%%%%%%%%%%%%%%%%%%%%%%%%%%%%%%%%%%%%%%%%%%%%%%%%

%__________________________________________________________________
%
% Table 4 - Optical / near infrared disk parameter ratios.
%
%__________________________________________________________________
%

\tabcolsep1.2mm

  \begin{table}[t]
  \caption[ ]{Optical/near infrared disk parameter ratios.  \\
   columns: (1) Sample used for the statistics; (2) -- (4) Ratio ($R/K$) of disk
   cut-off radii $\Rmax$, scale lengths $h$, and scale heights $\zo$, respectively;
   (5) Number of galaxies in the (sub-) sample.}
  \label{colours}
  \begin{flushleft}
  \begin{tabular}{llllc}
  \cline{1-5}
  \hline\hline\noalign{\smallskip}
  \multicolumn{1}{c}{Sample} & \multicolumn{3}{c}{Mean Ratio}  & \multicolumn{1}{c}{Number of} \\
  \noalign{\smallskip}
  \cline{2-4}
  \noalign{\smallskip}
  \multicolumn{1}{c}{}  & \multicolumn{1}{c}{$(\Rmax)_{R/K}$}  & \multicolumn{1}{c}{$h_{R/K}$} &
  \multicolumn{1}{c}{$(\zo)_{R/K}$} & \multicolumn{1}{c}{Galaxies}  \\
  \noalign{\smallskip}
  \multicolumn{1}{c}{(1)} & \multicolumn{1}{c}{(2)} & \multicolumn{1}{c}{(3)} &
  \multicolumn{1}{c}{(4)} & \multicolumn{1}{c}{(5)} \\
  \noalign{\smallskip}
  \hline\noalign{\smallskip}
   total        &  1.32 $\pm$ 0.4  &  1.26 $\pm$ 0.5  &  1.33 $\pm$ 0.4  &  41  \\
   non-int.     &  1.18 $\pm$ 0.5  &  1.25 $\pm$ 0.6  &  1.25 $\pm$ 0.5  &  20  \\
   int./merg.   &  1.47 $\pm$ 0.3  &  1.26 $\pm$ 0.8  &  1.40 $\pm$ 0.3  &  21  \\
  \noalign{\smallskip}
  \hline\noalign{\smallskip}
  \end{tabular}
  \end{flushleft}
  \end{table}

%%%%%%%%%%%%%%%%%%%%%%%%%%%%%%%%%%%%%%%%%%%%%%%%%%%%%%%%%%%%%%%%%%%

\subsection{Disk colour gradients}

In order to analyze colour gradients derived from measurements of radial and vertical
disk parameters in optical and in near infrared passbands ($R/K$), the mean ratios of
disk cut-off radius $\Rmax$, scale length $h$, and scale height $\zo$ -- obtained for
both subsamples -- are listed in Table~\ref{colours}.

The radial and vertical disk parameters found for $K$ and $R$ passbands in the {\em total}
galaxy sample indicate that the $R$-band values are systematically larger, i.e. of the
order of $(R/K) = 1.30 \pm 0.4$. In spite of the large intrinsic scatter, comparison with
literature data shows that the values (and also the errors) obtained here are consistent
with gradients derived by Giovanardi \& Hunt (\cite{giovanardi1988}) and de Grijs (\cite{grijs1997}).
They found $(F/K) = 1.20 \pm 0.42$ for the $F$- and $K$ passbands, and $(B/K) = 1.56 \pm 0.45$;
$(I/K) = 1.19 \pm 0.17$ for the $B$,- $I$- and $K$ passbands, respectively.

While systematically higher optical values have been found for each of the subsamples of
interacting/merging $(R/K) = 1.23 \pm 0.5$ and non-interacting galaxies $(R/K) = 1.38 \pm 0.5$,
the gradients of the latter sample are, however, systematically higher (Table~\ref{colours}).

Although the results obtained for the two subsamples are difficult to interpret, the systematic
differences found for all colour gradients as well as the good agreement with other studies
indicate that these gradients are not due to the large intrinsic errors. It should, however,
be stressed that the low S/N ratio in the outskirts of disks of a number of faint galaxies
obtained in the near infrared largely prevents a precise determination of the cut-off radius $\Rmax$.
Longer intergration times than the typical 30--40 min (on source) would be therefore necessary in
order to derive more reliable values.

%__________________________________________________________________

\section{Discussion}

Considering the small mass ratio between merging satellites and disks investigated here --
$M_{\rm sat}/M_{\rm disk} \approx 0.1$ -- the factor found for vertical disk thickening 
($\approx 1.6$ on average) and thus the efficiency of vertical heating is substantial.
This value is, however, significantly
lower than the factor of 2--4 obtained in previous studies (Reshetnikov \& Combes
\cite{reshetnikov1996}, \cite{reshetnikov1997}; Toth \& Ostriker \cite{toth1992}).
The differences can be explained by the different mass ratio between strongly interacting
systems (galaxies of comparable mass) investigated by Reshetnikov \& Combes, their
simplyfied disk model (isothermal) applied to all disks and the neglect of precise disk
inclination. In contrast to the set of fully self-consistent N-body simulations made recently
by Velazquez \& White (\cite{velazquez1999}) the analysis of Toth \& Ostriker (\cite{toth1992})
ignores the coherent response of the disk and its interaction with the halo. Additionally,
their assumption that the orbital energy of the satellite is deposited locally in the disk
is clearly unrealistic.

\medskip

In fact, the increase of disk scale height by a factor of $\approx 1.6$ found in this
study corresponds quite well with the value of 1.5--2 obtained by Velazquez \& White
(\cite{velazquez1999}).
However, as already mentioned by Velazquez \& White (\cite{velazquez1999}) and in the introduction
of this study, vertical disk thickening due to a minor merger crucially depends on many other factors
such as the mass and density profile of the sinking satellite, its orbit (prograde or retrograde),
the content of gas deposited in the disk, and -- presumably most important -- on the morphological
type of the galaxy. That, in turn, implies that tidally-triggered disk thickening strongly depends
on the B/D ratio and hence on the initial disk thickness of the galaxy, which is characterized by
the ratio $h/\zo$. This is confirmed by the $h/\zo$-statistics obtained in this study (Fig.~6), showing
a total lack of thin interacting/merging galaxies with $h/\zo > 7$. This result has some direct,
important consequences on the evolution of disk galaxies on cosmological timescales (Toomre
\cite{toomre1977}; Weil et al. \cite{weil1998}) and would also constrain the different scenarious
discussed for disk heating (Jenkins \& Binney \cite{jenkins1990}, \cite{jenkins1991}; Sanchez-Salcedo
\cite{sanchez1999}; Valluri \cite{valluri1993}). Therefore a more detailed study of the parameter
space using our supplementary N-body simulations, combined with the obtained results,
will be given in a forthcoming paper. The mentioned good quantitative agreement between simulation and
observation, however, indicates that -- in spite of the rather simple approach and the number of open
questions on the details of minor merger processes -- the changes of the vertical disk structure must
be of this order (Kleinschmidt et al. \cite{kleinschmidt1999}; Schwarzkopf \cite{schwarzkopf1999};
Schwarzkopf \& Dettmar \cite{schwarzkopf1998}, \cite{schwarzkopf1999_1}), i.e. somewhat lower
than previously argued.

Furthermore, it is still an open question whether the radial disk structure of galaxies suffering
interactions/minor mergers within the mass range studied here changes significantly. The differences
between both the disk cut-off and scale length statistics obtained (10\% and 20\% on average, resp.)
are just on the level of statistical significance and do therefore not allow for an interpretation.
On the other side, there is observational evidence that radial disk shrinking is a typical
aftermath of tidal interactions between galaxies with comparable masses (Reshetnikov \& Combes
\cite{reshetnikov1996}, \cite{reshetnikov1997}). If so, this should also apply to smaller
interactions and, in particular, to minor mergers (Schwarzkopf \cite{schwarzkopf1999}). For further
clarity on this point it is therefore necessary to analyze the radial behaviour of such galactic
disks in greater detail and on the basis of an expanded galaxy sample, preferably supported by
N-body simulations.

The fact that nearly all galactic disks investigated (93\%) possess vertical luminosity profiles
which are more sharply peaked than an isothermal distribution reinforces the results of previous
observational studies (e.g. de Grijs et al. \cite{grijsetal1997}; Schwarzkopf \& Dettmar
\cite{schwarzkopf1997}) that have ruled out the validity of the $\iso$-distribution as an
adequate quantitative description for most galactic disks, especially close to their plane.
This fact, together with the result that almost the same percentage of interacting
and non-interacting galaxies shows an identical vertical disk structure (with differences smaller
than 2\%, see Table~\ref{profiles}), indicates that regional damaging effects, asymmetries or
perturbations evoked by tidal interactions are non-persistent phenomena with lifetimes significantly
shorter than disk thickening. Furthermore, it implies that interactions/minor mergers within the
investigated mass range are not capable to destroy the initial vertical disk structure.

%__________________________________________________________________

\section{Summary and conclusions}

In this work a detailed statistical study is presented in order to investigate the effects
of minor mergers and tidal interactions in the range $M_{\rm sat}/M_{\rm disk} \approx 0.1$
on the radial and vertical structure of galactic disks.
The fundamental disk parameters of 110 highly-inclined/edge-on disk galaxies are determined
in optical and in near infrared passbands. This sample consists of two subsamples of 61
non-interacting and 49 strongly interacting/merging galaxies, respectively. Additionally,
41 of these galaxies were observed in the near infrared.
The main conclusions can be summarized as follows:

\begin{enumerate}

\item The structural changes of galactic disks affected by interaction/low-mass satellite
infall are most noticeable in the direction perpendicular to the disk plane.

\smallskip

\item While the majority of non-interacting galaxies possess a typical exponential disk scale
height of $\zo \approx 700$ pc, disks of minor mergers were found to be systematically thicker
with $\zo \approx 1.3$ kpc.

\smallskip

\item On average, galactic disks affected by interactions or minor mergers have $\approx~1.5$
times larger scale heights and thus vertical velocity dispersions than unperturbed disks.

\smallskip

\item The ratio of radial to vertical scale parameters, i.e. the normalized disk thickness
$h/\zo$, is $\approx~1.7$ times smaller for the interacting/merging sample.

\smallskip

\item Ratios $h/\zo > 7$, as they are typically for flat galaxies, are completely missing
in the interacting/merging sample. This implies that vertical disk heating is most efficient
for such (extremely) thin disks.

\smallskip

\item The radial disk structure of interacting/merging galaxies, characterized by the cut-off
radius $\Rmax$ and scale length $h$, shows no statistically significant changes.

\smallskip

\item The vertical luminosity profiles of all galactic disks investigated show the following
distribution (independent of the sample and passband): \newline
60\% $\propto$ sech; 33\% $\propto$ exp; 7\% $\propto \iso$.

\smallskip

\item Thus, the majority (93\%) of galactic disks possess non-isothermal vertical luminosity
profiles and is somewhat closer to a sech- than to an exp distribution.

\smallskip

\item There are no fundamental differences between the vertical luminosity distribution of
non-interacting and interacting/merging galaxies. Hence, the intrinsic distribution of disk
stars keeps largely retained during and after interactions/minor mergers.

\end{enumerate}

%__________________________________________________________________

\begin{acknowledgements}

We thank the referee of this series of papers, Dr. H. Wozniak, for his
useful comments and suggestions.
This work was supported by {\em Deutsche Forschungsgemeinschaft}, DFG,
under grant no. GRK118/2.
This research has made use of the NASA/IPAC Extragalactic Database (NED).

\end{acknowledgements}

%__________________________________________________________________

%\pagebreak

%__________________________________________________________________
%
% Table 5 - Global parameters and disk properties of the
%           optical / near infrared galaxy samples.
%
%__________________________________________________________________
%

\tabcolsep2.1mm

\begin{center}
  \begin{table*}
  \label{parameters}
  \caption[ ]{Global parameters and disk properties of the optical/near infrared galaxy samples. \\
   columns: (1) Serial number; (2) Galaxy name; (3) Passband; (4) Revised Hubble type, based on
   NASA/IPAC Extragalactic Database (NED); (5) Heliocentric velocity (if available), based on NED;
   (6) Galaxy distances, corrected for Virgocentric Flow (Sect. 2.2), using
   $H_{0} = 75 \, \rm km \, \rm s^{-1} Mpc^{-1}$; (7) Disk inclination; (8) Cut-off radius;
   (9) -- (10) Radial and vertical disk parameters, in case of two-components fits both parameters
   are given; (11) Best-fitting vertical disk model: $ 1= \rm exp, 2= \rm sech, 3= \iso$.}
  \begin{flushleft}
  \begin{tabular}{rlcrrrcrrcrrcrrc}
  \cline{1-16}
  \hline\hline
  \noalign{\smallskip}
  \multicolumn{1}{r}{No.}  & \multicolumn{1}{c}{Galaxy}  &   \multicolumn{1}{c}{Band} &
  \multicolumn{1}{r}{Type} & \multicolumn{1}{c}{$v_0$}   &   \multicolumn{1}{c}{Dist} &
  \multicolumn{1}{c}{$i$}  & \multicolumn{2}{c}{\Rmax}   & & \multicolumn{2}{c}{$h$}  & &
  \multicolumn{2}{c}{\zo}  &  $f(z)$   \\
  \noalign{\smallskip}
  \cline{8-9}
  \cline{11-12}
  \cline{14-15}
  \noalign{\smallskip}
   &  &  &  & [km/s] & [Mpc] & [\degr] & \multicolumn{1}{c}{[$''$]} & [kpc]
   &  & \multicolumn{1}{c}{[$''$]} & [kpc] &  & \multicolumn{1}{c}{[$''$]} & [kpc] & \\
  \noalign{\smallskip}
  \multicolumn{1}{r}{(1)}   & \multicolumn{1}{c}{(2)}    &   \multicolumn{1}{c}{(3)} &
  \multicolumn{1}{r}{(4)}   & \multicolumn{1}{c}{(5)}    &   \multicolumn{1}{c}{(6)} &
  \multicolumn{1}{c}{(7)}   & \multicolumn{2}{c}{(8)}    & & \multicolumn{2}{c}{(9)} & &
  \multicolumn{2}{c}{(10)}  & \multicolumn{1}{c}{(11)}   \\
  \noalign{\smallskip}
  \hline
%
%--------------------------------------------------------------------------------------------------------
% (1)   (2)         (3)    (4)     (5)    (6)     (7)        (8)              (9)           (10)     (11)
%
% No.   NAME        Band   Type    Vo     Dist     i         Rmax              h             zo      f(z)
%        
%                                                 degr   arcsec   kpc    arcsec   kpc    arcsec  kpc
%---------- merger --------------------------------------------------------------------------------------
%
  \noalign{\medskip}
  \multicolumn{16}{c}{\bf Interacting / Merging} \\
  \noalign{\medskip}
  \hline
  \noalign{\medskip}
%--------------------------------------------------------------------------------------------------------
  1 & NGC 7       & $R$ &  4.5 &  1497 &  19.8 & 90.0 &  70.9 &  6.81 &&  28.8 &  2.76 &&  9.2 & 0.88 & 2 \\
  2 & UGC 260     & $R$ &  6.0 &  2272 &  31.8 & 88.5 &  69.3 & 10.68 &&  20.6 &  3.18 &&  6.0 & 0.93 & 2 \\
  3 & NGC 128     & $R$ & -2.0 &  4241 &  59.7 & 90.0 &  85.5 & 24.75 &&  17.1 &  4.95 && 12.1 & 3.49 & 1 \\
    &             & $K$ & -2.0 &  4241 &  59.7 & 90.0 &  56.3 & 16.30 &&  16.0 &  4.63 &&  7.4 & 2.15 & 2 \\
  4 & AM 0107-375 & $R$ &  3.5 &  ---  &  ---  & 90.0 &  24.3 &  ---  &&   6.9 &  ---  &&  2.7 & ---  & 2 \\
  5 & ESO 296-G17 & $R$ &  3.0 &  5992 &  83.3 & 89.5 &  53.6 & 21.63 &&  18.8 &  7.61 &&  6.4 & 2.58 & 2 \\
  6 & ESO 354-G05 & $R$ &  4.0 &  ---  &  ---  & 90.0 &  30.7 &  ---  &&   7.9 &  ---  &&  2.8 & ---  & 2 \\
  7 & ESO 245-G10 & $R$ &  3.0 &  5713 &  79.6 & 89.0 &  46.1 & 17.80 &&  22.3 &  8.61 &&  7.9 & 3.06 & 2 \\
  8 & ESO 417-G08 & $R$ &  0.7 &  4893 &  67.5 & 89.5 &  64.5 & 21.10 &&  16.9 &  5.52 &&  7.9 & 2.60 & 2 \\
  9 & ESO 199-G12 & $R$ &  8.0 &  6993 &  96.5 & 90.0 &  40.7 & 19.03 &&  15.9 &  7.44 &&  3.6 & 1.71 & 2 \\
 10 & ESO 357-G16 & $R$ &  3.0 &  1382 &  18.2 & 90.0 &  54.5 &  4.81 &&  16.8 &  1.49 &&  5.4 & 0.48 & 2 \\
%
%--------------------------------------------------------------------------------------------------------
%
 11 & ESO 357-G26 & $R$ & -1.0 &  1362 &  18.3 & 89.6 &  79.4 &  7.04 &&  38.2 &  3.39 && 16.2 & 1.44 & 1 \\
 12 & ESO 418-G15 & $R$ &  4.0 &  1068 &  13.9 & 90.0 & 105.2 &  7.09 &&  34.7 &  2.34 &&  9.7 & 0.65 & 2 \\
 13 & NGC 1531/32 & $r$ &  2.7 &  1190 &  16.0 & 90.0 & 308.0 & 23.89 &&  96.0 &  7.45 && 38.0 & 2.94 & 2 \\
 14 & ESO 202-G04 & $R$ &  2.0 &   990 &  14.7 & 90.0 &  40.7 &  2.90 &&   8.4 &  0.60 &&  7.2 & 0.51 & 2 \\
 15 & ESO 362-G11 & $R$ &  4.0 &  1348 &  19.0 & 88.0 & 104.2 &  9.60 &&  31.7 &  2.92 && 13.5 & 1.24 & 1 \\
 16 & NGC 1888    & $r$ &  5.0 &  2432 &  31.8 & 90.0 &  96.8 & 14.92 &&  25.6 &  3.95 && 12.3 & 1.89 & 2 \\
    &             & $K'$&  5.0 &  2432 &  31.8 & 90.0 &  62.3 &  9.60 &&  14.9 &  2.29 &&  8.3 & 1.27 & 2 \\
 17 & ESO 363-G07 & $R$ &  5.5 &  1324 &  18.9 & 89.0 &  96.7 &  8.86 &&  32.7 &  3.00 && 10.7 & 0.98 & 1 \\
 18 & ESO 487-G35 & $R$ &  7.8 &  1731 &  23.6 & 90.0 &  74.4 &  8.51 &&  27.8 &  3.18 &&  9.7 & 1.11 & 2 \\
 19 & NGC 2188    & $r$ &  9.0 &   749 &  11.6 & 90.0 & 183.2 & 10.30 &&  35.2 &  1.98 && 20.3 & 1.14 & 1 \\
    &             & $K'$&  9.0 &   749 &  11.6 & 90.0 &  79.3 &  4.46 &&  38.9 &  2.19 && 13.3 & 0.75 & 2 \\
 20 & UGC 3697    & $R$ &  7.0 &  3137 &  43.5 & 90.0 &  65.2 & 13.75 &&  29.0 &  6.11 &&  3.0 & 0.62 & 2 \\
    &             & $H$ &  7.0 &  3137 &  43.5 & 90.0 &  73.2 & 15.44 &&  22.2 &  4.68 &&  5.7 & 1.21 & 2 \\
%
%--------------------------------------------------------------------------------------------------------
%
 21 & ESO 060-G24 & $r$ &  2.5 &  4096 &  57.6 & 87.0 &  97.5 & 27.23 &&  19.9 &  5.55 &&  6.8 & 1.89 & 2 \\
 22 & ESO 497-G14 & $R$ &  3.0 &  3446 &  45.4 & 90.0 &  45.6 & 10.04 &&  17.4 &  3.82 &&  5.9 & 1.29 & 2 \\
 23 & NGC 2820    & $H$ &  5.0 &  1580 &  31.2 & 90.0 &  78.0 & 11.80 &&  19.2 &  2.90 &&  7.8 & 1.18 & 2 \\
 24 & NGC 3044    & $r$ &  5.0 &  1292 &  21.4 & 89.0 & 179.2 & 18.59 &&  40.0 &  4.15 && 10.6 & 1.10 & 1 \\
    &             & $K'$&  5.0 &  1292 &  21.4 & 89.0 &  89.9 &  9.33 &&  27.6 &  2.86 &&  7.9 & 0.82 & 2 \\
 25 & NGC 3187    & $r$ &  5.0 &  1578 &  25.0 & 86.0 & 148.0 & 17.94 &&  26.0 &  3.15 && 27.9 & 3.39 & 1 \\
    &             & $K'$&  5.0 &  1578 &  25.0 & 86.0 &  84.2 & 10.21 &&  62.3 &  7.55 && 18.3 & 2.22 & 2 \\
 26 & ESO 317-G29 & $R$ &  1.0 &  2520 &  34.6 & 90.0 &  71.9 & 12.06 &&  24.8 &  4.16 && 11.5 & 1.93 & 2 \\
 27 & ESO 264-G29 & $R$ &  5.6 &  3310 &  45.6 & 90.0 &  42.7 &  9.43 &&  13.9 &  3.07 &&  4.3 & 0.94 & 2 \\
    &             & $K'$&  5.6 &  3310 &  45.6 & 90.0 &  38.2 &  8.45 &&   8.5 &  1.88 &&  2.5 & 0.55 & 2 \\
 28 & NGC 3432    & $R$ &  9.0 &   616 &  10.5 & 90.0 & 119.6 &  6.09 &&  29.0 &  1.48 && 14.3 & 0.73 & 1 \\
    &             & $K'$&  9.0 &   616 &  10.5 & 90.0 & 112.3 &  5.72 &&  64.2 &  3.27 && 16.7 & 0.85 & 1 \\
 29 & NGC 3628    & $r$ &  3.0 &   847 &   4.8 & 88.0 & 432.0 & 10.05 && 400.0 &  9.31 && 68.0 & 1.58 & 2 \\
    &             & $K'$&  3.0 &   847 &   4.8 & 88.0 & 231.0 &  5.38 &&  96.2 &  2.24 && 31.4 & 0.73 & 1 \\
 30 & ESO 378-G13 & $R$ &  1.0 &   --- &  ---  & 90.0 &  32.2 &  ---  &&  19.8 &  ---  &&  4.3 &  --- & 2 \\
    &             & $K'$&  1.0 &   --- &  ---  & 90.0 &  31.9 &  ---  &&  29.7 &  ---  &&  3.3 &  --- & 3 \\
%
%--------------------------------------------------------------------------------------------------------
%
 31 & ESO 379-G20 & $R$ &  1.0 & 14400 & 193.1 & 90.0 &  48.6 & 45.48 &&  14.9 & 13.92 &&  3.1 & 2.94 & 2 \\
    &             & $K'$&  1.0 & 14400 & 193.1 & 90.0 &  44.6 & 41.73 &&  12.7 & 11.92 &&  3.3 & 3.12 & 2 \\
 32 & NGC 4183    & $R$ &  6.0 &   932 &  17.6 & 90.0 & 137.8 & 11.76 &&  39.9 &  3.40 &&  8.7 & 0.74 & 2 \\
    &             & $R$ &  6.0 &   932 &  17.6 & 90.0 &  78.3 &  6.68 &&  39.9 &  3.40 &&  7.6 & 0.65 & 2 \\
    &             & $K'$&  6.0 &   932 &  17.6 & 90.0 &  88.5 &  7.55 &&  38.5 &  3.28 && 12.7 & 1.08 & 1 \\
 33 & NGC 4631    & $R$ &  7.0 &   606 &   8.0 & 89.0 & 261.0 & 10.12 &&  79.8 &  3.09 && 27.4 & 1.06 & 1 \\
    &             & $K'$&  7.0 &   606 &   8.0 & 89.0 & 241.0 &  9.33 &&  54.5 &  2.11 && 22.5 & 0.87 & 2 \\
%
%--------------------------------------------------------------------------------------------------------
%
  \noalign{\smallskip}
  \hline
  \end{tabular}
  \end{flushleft}
  \end{table*}
\end{center}

\tabcolsep2.0mm

\begin{center}
  \begin{table*}
 {\bf Table 5.} continued. \\
  \begin{flushleft}
  \begin{tabular}{rlcrrrcrrcrrcrrc}
  \cline{1-16}
  \hline\hline
  \noalign{\smallskip}
  \multicolumn{1}{r}{No.}  & \multicolumn{1}{c}{Galaxy}  &   \multicolumn{1}{c}{Band} &
  \multicolumn{1}{r}{Type} & \multicolumn{1}{c}{$v_0$}   &   \multicolumn{1}{c}{Dist} &
  \multicolumn{1}{c}{$i$}  & \multicolumn{2}{c}{\Rmax}   & & \multicolumn{2}{c}{$h$}  & &
  \multicolumn{2}{c}{\zo}  &  $f(z)$   \\
  \noalign{\smallskip}
  \cline{8-9}
  \cline{11-12}
  \cline{14-15}
  \noalign{\smallskip}
   &  &  &  & [km/s] & [Mpc] & [\degr] & \multicolumn{1}{c}{[$''$]} & [kpc]
   &  & \multicolumn{1}{c}{[$''$]} & [kpc] &  & \multicolumn{1}{c}{[$''$]} & [kpc] & \\
  \noalign{\smallskip}
  \multicolumn{1}{r}{(1)}   & \multicolumn{1}{c}{(2)}    &   \multicolumn{1}{c}{(3)} &
  \multicolumn{1}{r}{(4)}   & \multicolumn{1}{c}{(5)}    &   \multicolumn{1}{c}{(6)} &
  \multicolumn{1}{c}{(7)}   & \multicolumn{2}{c}{(8)}    & & \multicolumn{2}{c}{(9)} & &
  \multicolumn{2}{c}{(10)}  & \multicolumn{1}{c}{(11)}   \\
  \noalign{\smallskip}
  \hline
  \noalign{\medskip}
%
%--------------------------------------------------------------------------------------------------------
% (1)   (2)         (3)    (4)     (5)    (6)     (7)        (8)              (9)           (10)     (11)
%
% No.   NAME        Band   Type    Vo     Dist     i         Rmax              h             zo      f(z)
%        
%                                                 degr  arcsec   kpc     arcsec   kpc    arcsec  kpc
%--------------------------------------------------------------------------------------------------------
%
 34 & NGC 4634    & $r$ &  6.0 &   160 &  15.8 & 89.5 &  83.8 &  6.42 &&  27.3 &  2.09 &&  8.8 & 0.67 & 1 \\
    &             & $K$ &  6.0 &   160 &  15.8 & 89.5 &  65.0 &  4.98 &&  16.5 &  1.26 &&  6.0 & 0.46 & 1 \\
 35 & NGC 4747    & $R$ &  6.0 &  1188 &  22.4 & 88.0 &  79.8 &  8.66 &&  20.3 &  2.20 && 13.6 & 1.47 & 1 \\
 36 & NGC 4762    & $r$ & -2.0 &   985 &  11.7 & 87.0 & 272.0 & 15.43 && 164.0 &  9.30 && 23.1 & 1.31 & 2 \\
    &             & $r$ & -2.0 &   985 &  11.7 & 89.0 & 118.0 &  6.69 && 164.0 &  9.30 && 15.1 & 0.86 & 1 \\
    &             & $K'$& -2.0 &   985 &  11.7 & 89.0 & 160.4 &  9.10 && 128.3 &  7.28 && 11.4 & 0.65 & 3 \\
    &             & $K'$& -2.0 &   985 &  11.7 & 89.0 & 119.3 &  6.77 && 128.3 &  7.28 && 12.0 & 0.68 & 1 \\
 37 & ESO 443-G21 & $r$ &  6.0 &  2798 &  38.3 & 90.0 &  72.0 & 13.37 &&  16.0 &  2.97 &&  5.3 & 0.98 & 2 \\
    &             & $K'$&  6.0 &  2798 &  38.3 & 90.0 &  43.9 &  8.15 &&  22.7 &  4.21 &&  4.0 & 0.74 & 2 \\
 38 & NGC 5126    & $R$ &  0.0 &  4727 &  64.2 & 89.0 &  40.7 & 12.66 &&  14.9 &  4.63 &&  6.6 & 2.07 & 2 \\
 39 & ESO 324-G23 & $r$ &  6.5 &  1443 &  26.7 & 89.5 & 103.2 & 13.36 &&  31.2 &  4.04 && 16.7 & 2.16 & 1 \\
    &             & $K'$&  6.5 &  1443 &  26.7 & 89.5 &  74.2 &  9.60 &&  35.0 &  4.53 &&  8.5 & 1.10 & 2 \\
 40 & ESO 383-G05 & $r$ &  3.7 &  3626 &  50.0 & 89.0 & 108.4 & 26.28 &&  33.9 &  8.22 && 13.5 & 3.28 & 1 \\
%
%--------------------------------------------------------------------------------------------------------
%
 41 & NGC 5297    & $R$ &  4.5 &  2407 &  37.9 & 90.0 &  61.6 & 11.32 &&  19.6 &  3.60 && 10.2 & 1.88 & 2 \\
 42 & ESO 445-G63 & $r$ &  5.3 &  ---  &  ---  & 90.0 &  52.6 &  ---  &&  10.1 &  ---  &&  6.4 &  --- & 2 \\
 43 & NGC 5529    & $R$ &  5.0 &  2883 &  37.0 & 87.5 & 132.0 & 23.68 &&  23.2 &  4.16 &&  6.8 & 1.22 & 1 \\
    &             & $H$ &  5.0 &  2883 &  37.0 & 87.5 & 114.0 & 20.45 &&  48.0 &  8.61 &&  8.0 & 1.44 & 2 \\
 44 & NGC 5965    & $R$ &  3.0 &  3412 &  45.8 & 86.5 & 106.6 & 23.67 &&  27.6 &  6.12 &&  8.9 & 1.98 & 1 \\
 45 & NGC 6045    & $R$ &  5.0 & 10049 & 133.3 & 87.0 &  31.2 & 20.16 &&  15.1 &  9.75 &&  5.5 & 3.55 & 1 \\
    &             & $K$ &  5.0 & 10049 & 133.3 & 90.0 &  23.7 & 15.30 &&  22.4 & 14.48 &&  3.5 & 2.26 & 2 \\
 46 & NGC 6361    & $R$ &  3.0 &  3812 &  53.1 & 90.0 &  69.2 & 17.82 &&  21.1 &  5.44 && 11.0 & 2.83 & 1 \\
    &             &$nf\A$& 3.0 &  3812 &  53.1 & 90.0 &  80.8 & 20.79 &&  18.6 &  4.79 && 10.8 & 2.78 & 2 \\
    &             & $K$ &  3.0 &  3812 &  53.1 & 90.0 &  23.7 &  6.10 &&  22.4 &  5.77 &&  6.1 & 1.57 & 2 \\
 47 & Arp 121     & $R$ &  2.0 &  5489 &  75.1 & 90.0 &  42.8 & 15.57 &&  20.1 &  7.33 &&  5.8 & 2.10 & 2 \\
    &             & $K$ &  2.0 &  5489 &  75.1 & 90.0 &  32.0 & 11.65 &&  11.5 &  4.19 &&  4.4 & 1.61 & 2 \\
 48 & ESO 462-G07 & $K'$&  4.0 &  ---  &  ---  & 90.0 &  33.3 &  ---  &&  10.6 &  ---  &&  3.1 &  --- & 2 \\
 49 & IC 4991     & $r$ & -2.0 &  5660 &  79.9 & 90.0 &  39.2 & 15.18 &&  12.0 &  4.65 &&  2.5 & 0.97 & 3 \\
%
%--------------------------------------------------------------------------------------------------------
%
  \noalign{\medskip}
  \hline
  \noalign{\medskip}
  \multicolumn{16}{c}{\bf Non -- Interacting} \\
  \noalign{\medskip}
  \hline
  \noalign{\medskip}
%
%------------ superthin & andere ------------------------------------------------------------------------
%
  1 & UGC 231     & $R$ &  6.0 &   842 &  13.7 & 90.0 &  82.5 &  5.48 &&  39.6 &  2.63 &&  5.3 & 0.35 & 3 \\
    &             & $K$ &  6.0 &   842 &  13.7 & 90.0 &  62.7 &  4.17 &&  33.3 &  2.21 &&  6.1 & 0.41 & 3 \\
  2 & ESO 150-G07 & $r$ &  1.0 &   --- &  ---  & 90.0 &  29.5 &  ---  &&   6.9 &   --- &&  3.4 & ---  & 2 \\
% (!)
  3 & ESO 112-G04$\B$ &$r$&5.6 &   --- &  ---  & 87.5 &  45.0 &  ---  &&  22.7 &   --- &&  2.0 & ---  & 2 \\
  4 & ESO 150-G14$\B$ &$r$&0.4 &  8257 & 114.4 & 90.0 &  64.0 & 35.52 &&  23.3 & 12.93 &&  5.2 & 2.87 & 2 \\
  5 & UGC 711     & $R$ &  6.7 &  1979 &  27.0 & 90.0 &  85.3 & 11.17 &&  35.7 &  4.67 &&  6.2 & 0.82 & 2 \\
  6 & ESO 244-G48 & $r$ &  3.0 &   --- &  ---  & 87.0 &  45.3 &  ---  &&  13.5 &   --- &&  6.9 & ---  & 1 \\
% (!)
  7 & UGC 1839    & $R$ &  7.3 &  1536 &  20.7 & 90.0 &  75.4 &  7.57 &&  22.6 &  2.27 &&  6.8 & 0.69 & 2 \\
  8 & NGC 891     & $R$ &  3.0 &   528 &   9.5 & 89.4 & 365.4 & 16.83 && 165.0 &  7.60 && 23.6 & 1.09 & 2 \\
    &             & $K'$&  3.0 &   528 &   9.5 & 89.4 & 336.8 & 15.51 &&  99.4 &  4.58 && 14.9 & 0.69 & 1 \\
  9 & ESO 416-G25$\B$ &$r$&3.0 &  4997 &  68.9 & 87.0 &  69.8 & 23.32 &&  30.9 & 10.32 &&  4.8 & 1.61 & 2 \\
%   & UGC 2370    & $R$ &  unbrauchbar! \\
 10 & UGC 2411    & $R$ &  8.5 &  2547 &  37.3 & 90.0 &  90.8 & 16.42 &&  36.2 &  6.55 &&  4.5 & 0.82 & 2 \\
%
%--------------------------------------------------------------------------------------------------------
%
 11 & IC 1877     & $R$ &  3.0 &  ---  &  ---  & 90.0 &  19.7 &  ---  &&  12.2 &   --- &&  2.0 & ---  & 3 \\
% bei AM0302-504 = ao0012
 12 & ESO 201-G22 & $R$ &  5.0 &  4069 &  57.3 & 89.0 &  75.4 & 20.94 &&  31.2 &  8.68 &&  4.0 & 1.11 & 2 \\
 13 & NGC 1886    & $R$ &  3.5 &  1755 &  23.6 & 87.5 &  86.8 &  9.93 &&  20.8 &  2.38 &&  5.6 & 0.64 & 2 \\
    &             & $K'$&  3.5 &  1755 &  23.6 & 87.5 &  78.3 &  8.96 &&  22.5 &  2.57 &&  6.1 & 0.70 & 1 \\
 14 & UGC 3474    & $R$ &  6.0 &  3634 &  50.5 & 89.0 &  47.3 & 11.59 &&  23.2 &  5.68 &&  3.8 & 0.94 & 2 \\
    &             & $K'$&  6.0 &  3634 &  50.5 & 89.0 &  57.7 & 14.13 &&  20.4 &  5.00 &&  4.1 & 1.00 & 1 \\
 15 & NGC 2310    & $R$ & -2.0 &  1187 &  18.4 & 90.0 & 103.1 &  9.20 &&  26.3 &  2.35 && 10.2 & 0.91 & 2 \\
 16 & UGC 4278    & $R$ &  7.0 &   563 &  10.9 & 90.0 &  97.6 &  5.16 &&  29.0 &  1.53 &&  5.8 & 0.30 & 2 \\
    &             & $H$ &  7.0 &   563 &  10.9 & 90.0 & 168.0 &  8.88 &&  48.0 &  2.54 &&  9.9 & 0.52 & 2 \\
 17 & ESO 564-G27$\B$ &$r$&6.3 &  2178 &  33.2 & 87.0 & 131.4 & 21.15 &&  34.1 &  5.50 &&  4.0 & 0.64 & 2 \\
 18 & UGC 4943    & $R$ &  3.0 &  2265 &  34.2 & 90.0 &  50.2 &  8.33 &&  12.1 &  2.00 &&  1.7 & 0.29 & 2 \\
    &             & $H$ &  3.0 &  2265 &  34.2 & 90.0 &  49.2 &  8.16 &&  10.8 &  1.79 &&  2.9 & 0.48 & 2 \\
%
%--------------------------------------------------------------------------------------------------------
%
  \noalign{\smallskip}
  \hline
  \end{tabular}
  \end{flushleft}
  \end{table*}
\end{center}

\begin{center}
  \begin{table*}
 {\bf Table 5.} continued. \\
  \begin{flushleft}
  \begin{tabular}{rlcrrrcrrcrrcrrc}
  \cline{1-16}
  \hline\hline
  \noalign{\smallskip}
  \multicolumn{1}{r}{No.}  & \multicolumn{1}{c}{Galaxy}  &   \multicolumn{1}{c}{Band} &
  \multicolumn{1}{r}{Type} & \multicolumn{1}{c}{$v_0$}   &   \multicolumn{1}{c}{Dist} &
  \multicolumn{1}{c}{$i$}  & \multicolumn{2}{c}{\Rmax}   & & \multicolumn{2}{c}{$h$}  & &
  \multicolumn{2}{c}{\zo}  &  $f(z)$   \\
  \noalign{\smallskip}
  \cline{8-9}
  \cline{11-12}
  \cline{14-15}
  \noalign{\smallskip}
   &  &  &  & [km/s] & [Mpc] & [\degr] & \multicolumn{1}{c}{[$''$]} & [kpc]
   &  & \multicolumn{1}{c}{[$''$]} & [kpc] &  & \multicolumn{1}{c}{[$''$]} & [kpc] & \\
  \noalign{\smallskip}
  \multicolumn{1}{r}{(1)}   & \multicolumn{1}{c}{(2)}    &   \multicolumn{1}{c}{(3)} &
  \multicolumn{1}{r}{(4)}   & \multicolumn{1}{c}{(5)}    &   \multicolumn{1}{c}{(6)} &
  \multicolumn{1}{c}{(7)}   & \multicolumn{2}{c}{(8)}    & & \multicolumn{2}{c}{(9)} & &
  \multicolumn{2}{c}{(10)}  & \multicolumn{1}{c}{(11)}   \\
  \noalign{\smallskip}
  \hline\noalign{\medskip}
%
%--------------------------------------------------------------------------------------------------------
% (1)   (2)         (3)    (4)     (5)    (6)     (7)        (8)              (9)           (10)     (11)
%
% No.   NAME        Band   Type    Vo     Dist     i         Rmax              h             zo      f(z)
%        
%                                                 degr  arcsec   kpc     arcsec   kpc    arcsec  kpc
%--------------------------------------------------------------------------------------------------------
%
 19 & IC 2469     & $r$ &  2.0 &  1666 &  22.2 & 82.0 & 197.0 & 25.79 &&  45.2 &  5.92 && 13.9 & 1.50 & 1 \\
 20 & UGC 5341    & $R$ &  6.0 &  7568 &  97.3 & 89.0 &  86.8 & 40.95 &&  31.7 & 14.97 &&  4.2 & 2.00 & 2 \\
    &             & $K'$&  6.0 &  7568 &  97.3 & 89.0 &  66.1 & 31.17 &&  19.1 &  9.03 &&  2.7 & 1.26 & 1 \\
%
%--------------------------------------------------------------------------------------------------------
%
 21 & IC 2531     & $r$ &  5.3 &  2477 &  33.0 & 89.2 & 220.0 & 35.20 &&  67.2 & 10.75 &&  9.8 & 1.57 & 2 \\
% 23 & NGC 3115    & $r$ & -3.0 &   658 &  11.0 & 90.0 & 292.5 & 16.00 &&  50.7 &  2.70 && 64.3 & 3.43 & 1 \\
 22 & NGC 3390    & $r$ &  3.0 &  2820 &  37.9 & 89.0 & 113.1 & 20.78 &&  23.4 &  4.30 &&  9.2 & 1.69 & 1 \\
 23 & ESO 319-G26$\B$ &$r$&5.3 &  3601 &  42.4 & 89.5 &  51.8 & 10.65 &&  14.3 &  2.93 &&  2.0 & 0.41 & 3 \\
 24 & NGC 3957    & $r$ & -1.0 &  1703 &  29.6 & 89.0 &  93.6 & 13.43 &&  23.4 &  3.36 &&  8.4 & 1.20 & 1 \\
 25 & NGC 4013    & $R$ &  3.0 &   839 &  12.0 & 89.5 &  96.4 &  5.61 &&  24.6 &  1.43 && 11.6 & 0.67 & 1 \\
    &             & $K'$&  3.0 &   839 &  12.0 & 89.5 & 117.4 &  6.83 &&  43.6 &  2.54 &&  9.7 & 0.57 & 1 \\
 26 & ESO 572-G44 & $r$ &  3.0 &  6759 &  89.9 & 89.5 &  50.4 & 21.95 &&  33.1 & 14.41 &&  6.5 & 2.82 & 1 \\
% (!)
 27 & UGC 7170    & $R$ &  6.0 &  2444 &  29.7 & 90.0 &  54.1 &  7.79 &&  16.4 &  2.36 &&  2.8 & 0.41 & 2 \\
 28 & ESO 321-G10$\B$ &$r$&1.4 &  3147 &  42.4 & 88.0 &  64.0 & 13.17 &&  25.8 &  5.31 &&  4.8 & 0.99 & 1 \\
 29 & NGC 4217    & $R$ &  3.0 &  1026 &  14.6 & 89.0 & 150.8 & 10.67 &&  29.0 &  2.05 && 21.6 & 1.53 & 2 \\
 30 & NGC 4244    & $R$ &  6.0 &   243 &   3.8 & 90.0 & 322.6 &  5.94 &&  76.1 &  1.40 && 18.4 & 0.34 & 2 \\
%
%--------------------------------------------------------------------------------------------------------
%
 31 & UGC 7321    & $R$ &  7.0 &   409 &  15.8 & 90.0 & 103.4 &  7.92 &&  44.4 &  3.40 &&  4.7 & 0.36 & 2 \\
    &             & $H$ &  7.0 &   409 &  15.8 & 90.0 & 163.2 & 12.50 &&  43.2 &  3.31 &&  7.1 & 0.55 & 2 \\
 32 & NGC 4302    & $R$ &  5.0 &  1108 &  15.8 & 90.0 & 181.6 & 13.91 &&  74.2 &  5.68 &&  7.0 & 0.53 & 2 \\
    &             & $H$ &  5.0 &  1108 &  15.8 & 90.0 & 145.2 & 11.12 &&  66.0 &  5.06 && 13.0 & 0.99 & 2 \\
 33 & NGC 4330    & $r$ &  6.0 &  1569 &  15.8 & 88.5 & 290.0 & 22.21 && 108.8 &  8.33 && 25.3 & 1.94 & 1 \\
    &             & $K'$&  6.0 &  1569 &  15.8 & 88.5 &  95.6 &  7.32 &&  39.6 &  3.04 &&  7.9 & 0.60 & 2 \\
 34 & NGC 4565    & $R$ &  3.0 &  1227 &  10.0 & 88.0 & 229.4 & 11.12 &&  67.6 &  3.28 && 10.8 & 0.52 & 1 \\
    &             & $K'$&  3.0 &  1227 &  10.0 & 88.0 & 311.2 & 15.09 &&  96.2 &  4.66 && 15.3 & 0.74 & 2 \\
 35 & NGC 4710    & $R$ & -1.0 &  1119 &  14.4 & 87.0 & 134.1 &  9.36 &&  39.9 &  2.78 && 18.0 & 1.25 & 2 \\
    &             & $R$ & -1.0 &  1119 &  14.4 & 87.0 &  79.8 &  5.57 &&  16.7 &  1.16 &&  9.8 & 0.69 & 1 \\
    &             & $K'$& -1.0 &  1119 &  14.4 & 87.0 & 117.4 &  8.20 &&  38.5 &  2.69 && 12.4 & 0.86 & 2 \\
% 39 & NGC 5124    & $R$ & -5.0 &  3976 &  50.9 & 90.0 &  57.5 & 14.20 &&  15.9 &  3.92 && 11.2 & 2.77 & 2 \\
% bei NGC 5126 = ao0013 
 36 & NGC 5170    & $r$ &  5.0 &  1503 &  15.8 & 88.0 & 272.0 & 20.84 && 124.0 &  9.50 && 20.3 & 1.56 & 1 \\
    &             & $H$ &  5.0 &  1503 &  15.8 & 88.0 & 222.0 & 17.01 &&  96.0 &  7.35 && 15.4 & 1.18 & 2 \\
    &             & $K$ &  5.0 &  1503 &  15.8 & 88.0 & 222.0 & 17.01 &&  66.0 &  5.06 && 11.7 & 0.90 & 2 \\
 37 & ESO 510-G18 & $r$ &  1.0 &   --- &   --- & 90.0 &  16.4 &  ---  &&   3.8 &  ---  &&  1.2 & ---  & 2 \\
% (!)
 38 & UGC 9242    & $R$ &  7.0 &  1440 &  25.2 & 88.5 &  93.7 & 11.45 &&  41.1 &  5.02 &&  4.7 & 0.57 & 1 \\
    &             & $H$ &  7.0 &  1440 &  25.2 & 88.5 &  70.4 &  8.60 &&  19.2 &  2.35 &&  1.8 & 0.22 & 3 \\
 39 & NGC 5775    & $r$ &  5.0 &  1681 &  28.9 & 87.0 & 127.1 & 17.81 &&  31.2 &  4.37 && 13.1 & 1.84 & 1 \\
 40 & NGC 5907    & $R$ &  5.0 &   667 &  11.0 & 88.0 & 276.0 & 14.72 && 145.9 &  7.78 && 16.9 & 0.90 & 2 \\
    &             & $H$ &  5.0 &   667 &  11.0 & 88.0 & 225.6 & 12.03 &&  48.0 &  2.56 && 12.3 & 0.66 & 2 \\
%
%--------------------------------------------------------------------------------------------------------
%
 41 & NGC 5908    & $R$ &  3.0 &  3306 &  44.7 & 88.0 &  78.0 & 16.89 &&  20.1 &  4.36 &&  7.6 & 1.66 & 2 \\
    &             & $K'$&  3.0 &  3306 &  44.7 & 88.0 &  78.9 & 17.10 &&  24.3 &  5.26 &&  7.4 & 1.61 & 1 \\
 42 & ESO 583-G08 & $r$ &  4.0 &  7451 & 101.4 & 86.7 &  53.4 & 26.27 &&  12.5 &  6.14 &&  1.8 & 0.88 & 2 \\
 43 & NGC 6181    & $R$ &  5.0 &  2375 &  36.0 & 90.0 &  30.4 &  5.31 &&  13.0 &  2.28 &&  4.6 & 0.80 & 2 \\
 44 & ESO 230-G11 & $r$ &  4.0 &  5258 &  69.0 & 89.0 &  38.8 & 12.99 &&  15.0 &  5.01 &&  4.2 & 1.42 & 3 \\
% (!)
 45 & NGC 6722$\B$& $r$ &  3.0 &  4626 &  66.3 & 87.5 &  84.6 & 27.19 &&  24.0 &  7.73 &&  7.3 & 2.34 & 1 \\
 46 & ESO 461-G06 & $r$ &  5.0 &  ---  &  ---  & 88.8 &  60.8 &  ---  &&  16.0 &  ---  &&  3.0 &  --- & 1 \\
 47 & ESO 339-G16$\B$ &$r$&1.0 &  ---  &  ---  & 87.0 &  45.7 &  ---  &&  11.4 &  ---  &&  4.0 &  --- & 2 \\
 48 & IC 4937$\B$ & $r$ &  3.0 & $-$84 &  63.0 & 89.8 &  75.6 & 23.08 &&  27.3 &  8.33 &&  6.7 & 2.04 & 2 \\
 49 & ESO 187-G08 & $r$ &  6.0 &  4374 &  63.1 & 90.0 &  55.8 & 17.07 &&  14.8 &  4.52 &&  3.2 & 0.98 & 3 \\
 50 & IC 5052     & $r$ &  7.0 &   591 &   8.0 & 89.5 & 200.9 &  7.79 &&  48.8 &  1.89 && 19.8 & 0.77 & 1 \\
    &             & $K'$&  7.0 &   591 &   8.0 & 89.5 & 135.9 &  5.27 &&  67.3 &  2.61 && 11.9 & 0.46 & 2 \\
%
%--------------------------------------------------------------------------------------------------------
%
 51 & IC 5096     & $r$ &  4.0 &  3087 &  45.9 & 87.0 &  95.4 & 21.22 &&  32.3 &  7.20 &&  6.5 & 1.46 & 1 \\
% (!)
 52 & ESO 466-G01$\B$ &$r$&2.0 &  7095 &  98.7 & 88.0 &  56.1 & 26.86 &&  13.2 &  6.30 &&  7.7 & 3.67 & 1 \\
% (!)
 53 & ESO 189-G12 & $r$ &  5.0 &  8370 & 116.3 & 88.0 &  55.8 & 31.46 &&  23.0 & 12.99 &&  3.8 & 2.16 & 1 \\
% (!)
 54 & UGC 11859$\B$&$r$ &  4.0 &  3014 &  45.1 & 88.0 & 106.9 & 23.38 &&  25.2 &  5.51 &&  1.5 & 0.32 & 1 \\
 55 & ESO 533-G04 & $r$ &  4.8 &  2569 &  38.2 & 87.0 &  74.9 & 13.87 &&  25.2 &  4.67 &&  6.7 & 1.24 & 1 \\
 56 & IC 5199     & $r$ &  3.0 &  5037 &  71.5 & 86.5 &  58.5 & 20.28 &&  17.6 &  6.08 &&  4.5 & 1.57 & 1 \\
 57 & UGC 11994   & $R$ &  4.0 &  4872 &  70.6 & 90.0 &  70.1 & 24.00 &&  14.8 &  5.08 &&  2.3 & 0.80 & 3 \\
%
%--------------------------------------------------------------------------------------------------------
%
  \noalign{\smallskip}
  \hline
  \end{tabular}
  \end{flushleft}
  \end{table*}
\end{center}

\begin{center}
  \begin{table*}
 {\bf Table 5.} continued. \\
  \begin{flushleft}
  \begin{tabular}{rlcrrrcrrcrrcrrc}
  \cline{1-16}
  \hline\hline
  \noalign{\smallskip}
  \multicolumn{1}{r}{No.}  & \multicolumn{1}{c}{Galaxy}  &   \multicolumn{1}{c}{Band} &
  \multicolumn{1}{r}{Type} & \multicolumn{1}{c}{$v_0$}   &   \multicolumn{1}{c}{Dist} &
  \multicolumn{1}{c}{$i$}  & \multicolumn{2}{c}{\Rmax}   & & \multicolumn{2}{c}{$h$}  & &
  \multicolumn{2}{c}{\zo}  &  $f(z)$   \\
  \noalign{\smallskip}
  \cline{8-9}
  \cline{11-12}
  \cline{14-15}
  \noalign{\smallskip}
   &  &  &  & [km/s] & [Mpc] & [\degr] & \multicolumn{1}{c}{[$''$]} & [kpc]
   &  & \multicolumn{1}{c}{[$''$]} & [kpc] &  & \multicolumn{1}{c}{[$''$]} & [kpc] & \\
  \noalign{\smallskip}
  \multicolumn{1}{r}{(1)}   & \multicolumn{1}{c}{(2)}    &   \multicolumn{1}{c}{(3)} &
  \multicolumn{1}{r}{(4)}   & \multicolumn{1}{c}{(5)}    &   \multicolumn{1}{c}{(6)} &
  \multicolumn{1}{c}{(7)}   & \multicolumn{2}{c}{(8)}    & & \multicolumn{2}{c}{(9)} & &
  \multicolumn{2}{c}{(10)}  & \multicolumn{1}{c}{(11)}   \\
  \noalign{\smallskip}
  \hline\noalign{\medskip}
%
%--------------------------------------------------------------------------------------------------------
% (1)   (2)         (3)    (4)     (5)    (6)     (7)        (8)              (9)           (10)     (11)
%
% No.   NAME        Band   Type    Vo     Dist     i         Rmax              h             zo      f(z)
%        
%                                                 degr  arcsec   kpc     arcsec   kpc    arcsec  kpc
%--------------------------------------------------------------------------------------------------------
%
 58 & UGC 12281   & $R$ &  8.0 &  2567 &  39.5 & 90.0 & 103.1 & 19.75 &&  40.2 &  7.71 &&  6.0 & 1.16 & 2 \\
    &             & $K$ &  8.0 &  2567 &  39.5 & 90.0 &  56.3 & 10.79 &&  25.6 &  4.90 &&  5.1 & 0.98 & 1 \\
 59 & UGC 12423   & $R$ &  5.0 &  4839 &  70.0 & 87.4 & 100.6 & 34.14 &&  25.1 &  8.54 &&  6.6 & 2.23 & 1 \\
    &             & $K$ &  5.0 &  4839 &  70.0 & 87.4 &  53.8 & 18.24 &&  17.9 &  6.08 &&  4.0 & 1.70 & 2 \\
%
%--------------------------------------------------------------------------------------------------------
%
 60 & NGC 7518    & $R$ &  1.0 &  3536 &  52.6 & 87.0 &  55.1 & 14.05 &&  13.0 &  3.33 &&  7.6 & 1.93 & 1 \\
 61 & ESO 604-G06 & $r$ &  4.0 &  7665 & 106.1 & 89.6 &  74.9 & 38.52 &&  22.6 & 11.64 &&  3.0 & 1.54 & 2 \\
%
%--------------------------------------------------------------------------------------------------------
%
  \noalign{\smallskip}
  \hline
  \end{tabular}
  \begin{list}{}{}
  \item[$\A$] nf=no filter used. \\
  \item[$\B$] Supplementary data from Barteldrees \& Dettmar (\cite{barteldrees1994}).

\vspace{5mm}

  \end{list}
  \end{flushleft}
  \end{table*}
\end{center}

%__________________________________________________________________


\begin{thebibliography}{}

 \bibitem[1966]{arp1966}
  Arp H.C., 1966, Atlas of Peculiar Galaxies, California Institute of Technology,
  Pasadena

 \bibitem[1987]{arp1987}
  Arp H.C., Madore B., F., 1987, A Catalogue of Southern Peculiar Galaxies
  and Associations, Vol. I\&II, Cambridge University Press

 \bibitem[1992]{barnaby1992}
  Barnaby D., Thronson Jr. Harley A., 1992, AJ 103, 41

 \bibitem[1994]{barteldrees1994}
  Barteldrees A., Dettmar R.-J., 1994, A\&AS 103, 475

 \bibitem[1993]{bottema1993}
  Bottema R., 1993, A\&A 275, 16

 \bibitem[1996]{burkert1996}
  Burkert A., Yoshii Y., 1996, MNRAS 282, 1349

 \bibitem[1989]{carlberg1989}
  Carlberg R.G., Couchman H.M.P., 1989, ApJ 340, 47

 \bibitem[1957]{darling1957}
  Darling D.A., 1957, Ann. Math. Statist. 28, 823

 \bibitem[1988]{frenk1988}
  Frenk C.S., White S.D.M., Davis M., 1988, ApJ 327, 507

 \bibitem[1988]{fried1988}
  Fried J.W., 1988, A\&A 189, 42

 \bibitem[1990]{gerin1990}
  Gerin M., Combes F., Athanassoula E., 1990, A\&A 230, 37

 \bibitem[1983]{gilmore1983}
  Gilmore G., Reid N., 1983, MNRAS 202, 1025

 \bibitem[1989]{gilmore1989}
  Gilmore G., King I., Kruit P.C. van der, 1989, In: The Milky Way as a Galaxy.
  Geneva Observatory, Sauverny-Versoix, Switzerland

 \bibitem[1988]{giovanardi1988}
  Giovanardy C., Hunt L.K., 1988, AJ 95, 408

 \bibitem[1997]{grijs1997}
  Grijs R. de, 1997, PhD Thesis, University of Groningen

 \bibitem[1996]{grijs1996}
  Grijs R. de, Kruit P.C. van der, 1996, A\&AS 117, 19

 \bibitem[1997]{grijsetal1997}
  Grijs R. de, Pelletier R.F., Kruit P.C. van der, 1997, A\&A 327, 986

 \bibitem[1988]{habing1988}
  Habing H., 1988, A\&A 200, 40

 \bibitem[1998]{ibata1998}
  Ibata R.A., Lewis G.F, 1998, ApJ 500, 575

 \bibitem[1985]{irwin1985}
  Irwin M.J., Kunkel W.E., Demers S., 1985, Nature, 318, 160

 \bibitem[1990]{jenkins1990}
  Jenkins A., Binney J., 1990, MNRAS 245, 305

 \bibitem[1991]{jenkins1991}
  Jenkins A., Binney J., In: Dynamics of Galaxies and Their Molecular Cloud Distributions,
  Proceedings of the 146th IAU Symposium, Paris, France, eds. F. Combes \& Fabienne Casoli,
  Kluwer Publ., Dordrecht, 1991., p.~400

 \bibitem[1999]{kleinschmidt1999}
  Kleinschmidt L., Theis C., Schwarzkopf U., 1999, In: Astronomische Gesellschaft
  Abstract Series No. 15, 36

% \bibitem[1984]{kormendy1984}
%  Kormendy J., 1984, ApJ 287, 577

% \bibitem[1989]{kormendy1989}
%  Kormendy J., 1989, ApJ 342, L 63

 \bibitem[1986]{kraan-korteweg1986}
  Kraan-Korteweg R.C., 1986, A\&AS 66, 255

 \bibitem[1986]{kruit1986}
  Kruit P.C. van der, 1986, A\&A 157, 230

 \bibitem[1981a]{kruit1981a}
  Kruit P.C. van der, Searle L., 1981a, A\&A 95, 105

 \bibitem[b]{kruit1981b}
  Kruit P.C. van der, Searle L., 1981b, A\&A 95, 116

 \bibitem[1982a]{kruit1982a}
  Kruit P.C. van der, Searle L., 1982a, A\&A 110, 61

% \bibitem[b]{kruit1982b}
%  Kruit P.C. van der, Searle L., 1982b, A\&A 110, 79

 \bibitem[1995]{mihos1995}
  Mihos J.C., Walker I.R., Hernquist L., 1995, ApJ 447, L87

 \bibitem[1990]{ostriker1990}
  Ostriker J.P., 1990, In: Evolution of the Universe of Galaxies.
  ASP Conf. Ser. Vol. 10, p.23, ASP Publ., San Francisco

 \bibitem[1993]{quinn1993}
  Quinn P.J., Hernquist L., Fullagar D.P., 1993, ApJ 403, 74

 \bibitem[1996]{reshetnikov1996}
  Reshetnikov V.P., Combes F., 1996, A\&AS 116, 417

 \bibitem[1997]{reshetnikov1997}
  Reshetnikov V.P., Combes F., 1997, A\&A 324, 80

 \bibitem[1993]{reshetnikov1993}
  Reshetnikov V.P., Hagen-Thorn V.A., Yakovleva V.A., 1993, A\&A 278, 351

 \bibitem[1987]{richter1987}
  Richter O.-G., Tammann G.A., Huchtmeier W.K., 1987, A\&A 171, 33

 \bibitem[1992]{sachs1992}
  Sachs L., 1992, Angewandte Statistik, Springer-Verlag Berlin Heidelberg

 \bibitem[1999]{sanchez1999}
  Sanchez-Salcedo F. J.,, 1999, MNRAS 303, 755

 \bibitem[1992]{schommer1992}
  Schommer R.A., Olszewski E.W., Suntzeff N.B., Harris H.C., 1992, AJ 103, 447

 \bibitem[1999]{schwarzkopf1999}
  Schwarzkopf U., 1999, PhD Thesis, Ruhr-University Bochum

 \bibitem[1997]{schwarzkopf1997}
  Schwarzkopf U., Dettmar R.-J., 1997, In: Astronomische Gesellschaft Abstract
  Series No. 13, 238

 \bibitem[1998]{schwarzkopf1998}
  Schwarzkopf U., Dettmar R.-J., 1998, In: The Magellanic Clouds and Other Dwarf Galaxies.
  eds. Braun J.M., Richtler T., Shaker Publ., Aachen, p.~297

 \bibitem[1999]{schwarzkopf1999_1}
  Schwarzkopf U., Dettmar R.-J., 1999, Ap\&SS 265, 479

 \bibitem[2000a]{schwarzkopf2000a}
  Schwarzkopf U., Dettmar R.-J., 2000a, Paper~I, A\&AS 144, 85

 \bibitem[2000c]{schwarzkopf2000c}
  Schwarzkopf U., Dettmar R.-J., 2000c, Paper~III, A\&A, subm.

% \bibitem[1986]{schweitzer1986}
%  Schweitzer J., 1986, In: Nearly Normal Galaxies. p~18, ed. S.M. Faber, Springer-Verlag,
%  New York

 \bibitem[1989]{shaw1989}
  Shaw M.A., Gilmore G., 1989, MNRAS 237, 903

% \bibitem[1996]{skillman1996}
%  Skillman E.D., Kennicutt R.C., Shields G.A., Zaritsky D., 1996, ApJ 462, 147

 \bibitem[1977]{toomre1977}
  Toomre A., 1977, In: The evolution of galaxies and stellar populations, eds . B. Tinsley \&
  R. Larson, New Haven, Yale University Press, p.~401 

 \bibitem[1992]{toth1992}
  Toth G., Ostriker J.P., 1992, ApJ 389, 5

 \bibitem[1993]{valluri1993}
  Valluri M. 1993, ApJ 408, 57

 \bibitem[1999]{velazquez1999}
  Velazquez H., White S.D.M., 1999, MNRAS 304, 254

 \bibitem[1989]{wainscoat1989}
  Wainscoat R.J., Freeman K.C., Hyland A.R., 1989, ApJ 337, 163


\newpage


 \bibitem[1990]{wainscoat1990}
  Wainscoat R.J., Hyland A.R., Freeman K.C., 1990, ApJ 348, 85

 \bibitem[1996]{walker1996}
  Walker I.R., Mihos C., Hernquist L., 1996, ApJ 460, 121

 \bibitem[1998]{weil1998}
  Weil M. L., Eke V. R., Efstathiou G., 1998, MNRAS 300, 773

 \bibitem[1995]{zaritsky1995}
  Zaritsky D., 1995, ApJ 448, L17

% \bibitem[1996]{zaritsky1996}
%  Zaritsky D., 1996, ApJ 462, 50

 \bibitem[1993]{zaritsky1993}
  Zaritsky D., Smith R., Frenk C., White S.D.M., 1993, ApJ 405, 464

 \bibitem[1994]{zaritsky1994}
  Zaritsky D., Kennicutt R.C., Huchra J.P., 1994, ApJ 420, 87

 \bibitem[1997]{zaritsky1997}
  Zaritsky D., Smith R., Frenk C., White S.D.M., 1997, ApJ 478, 39


%\newpage % if processed with CM fonts, balances the references columns

\end{thebibliography}
\end{document}